\documentstyle[12pt,epsf]{article}
\author{Stefan
G\"ullenstern \\
{\normalsize Max-Planck-Institut for Nuclear Physics, P.O. Box 103 980} \\
{\normalsize 69029 Heidelberg, Germany}
\and Pawe\l\ G\'ornicki \\
{\normalsize  Institute of Physics and Center for Theoretical
Physics}\\
{\normalsize  Polish
Academy of Sciences}\\
{\normalsize Al. Lotnik\'ow 32/46, 02-668 Warsaw, Poland}\\
\and Lech Mankiewicz \\
{\normalsize Nicolaus Copernicus Astronomical Center, ul. Bartycka 18}\\
{\normalsize PL-00 716 Warsaw, Poland}\\
\and Andreas Sch\"afer \\
{\normalsize Institute for Theoretical Physics, University
Frankfurt}\\
{\normalsize D-60054 Frankfurt am Main, Germany}}
\title{Monte Carlo Simulations for RHIC Spin Physics}
\date{~~}
\begin{document}
\maketitle
\newpage
\abstract{
Direct photon production in longitudinally polarised
proton-proton collisions offers
the most direct and unproblematic possibility
to determine the polarised gluon distribution of a proton. This
information could play a major role for improving our understanding
of the nucleon structure and QCD in general. It is hoped that such
experiments will be done at RHIC. We present results of detailed Monte Carlo
simulations using a code called {\sc SPHINX}. We find that for RHIC
energies and large gluon polarisation the Compton graph dominates
allowing for a direct test of $\Delta g$.
Triggering on away-side jets with the envisaged jet-criteria
should allow to obtain more detailed information on $\Delta g(x)$.
The photon asymmetry resulting from the asymmetry of produced
$\pi^0$'s provides an additional signal, which is complementary to
the other two. For small gluon
polarisation, i.e. $\Delta g \le 0.5$ or very soft polarised
gluon-distributions the envisaged experiments will require a highly
sophisticated simulation and large statistics to extract
more than upper bounds for $|\Delta
g(x)|$.}
\vskip 4 cm
PACS-No.: ~~~~13.88.+e, 13.85.Hd, 12.38.Bx
\newpage
\section{Introduction}
The experiments on
longitudinally polarised lepton-nucleon scattering performed over
the last years at CERN \cite{EMC,SMC} and SLAC \cite{SLAC}
have determined the polarised structure function
$g_1^p(x)$ with rather good accuracy, while $g_1^n(x)$ still needs
improvement. The theoretical interpretation of these data is the
subject of intense debates which we do not want to review here.
These discussions have,
however, established a number of facts which imply the importance of
analysing polarised proton-proton collisions:\\
1.) Polarized reactions provide very sensitive tests of QCD. To
realize this potential fully requires, however, a detailed
experimental knowledge and theoretical understanding of
$Q^2$-dependences, just as for unpolarised reactions \cite{theo1}.
To disentangle the various effects precise and complementary
measurements will be needed. In addition comparing specific results
from deep-inelastic scattering and proton-proton collisions allows
for interesting tests. \cite{Ehrns}\\
2.) The polarised gluon distribution plays a very special role as it
contributes via the anomaly \cite{anoma}
in the same way as the polarised quark
distribution. For deep inelastic scattering there is no way to
distinguish between a `genuine' quark distribution and an `anomalous'
gluon distribution. Such a distinction is by principle
only possible by combining
inclusive and exclusive  data, see e.\ g.\ \cite{AS1} for polarised
deep-inelastic scattering, though probably very
difficult in praxis. On the other hand to establish the existence of
an anomalous gluon contribution would test fundamental topological
properties of QCD, which have never been directly accessible before.  \\
Obviously in this situation a precise knowledge of $\Delta g(x,Q^2)$
would help
tremendously, and this could be provided by polarised proton-proton
scattering. Such experiments would be possible at RHIC in the near
future,
details can be found in ref. \cite{RSC}.\\
Actually there are many interesting quantities which can be
determined in such experiments, also for transverse
polarisation, but their theoretical understanding is
at least partially incomplete and the
reliable simulation of background
reactions will require fundamentally modified codes. We therefore
concentrated on the measurement of $\Delta g$ via direct photon
assymmetries, which is the most straight forward experiment proposed
so far \cite{promptg}.    \\
Still it is generally accepted (and will be demonstrated in this
contribution) that planning and analysing such experiments requires a
full-fledged Monte-Carlo code. We developped one such code to
describe the collision of longitudinally polarised nucleons. It is
called {\sc SPHINX} and is basically a polarised version of
{\sc PYTHIA} \cite{pyman}. A description of this code will be published
elsewhere
\cite{SPHINX}, and we shall try to make  it generally accessible. In
our paper we present results obtained with this code for experiments
planned at RHIC using the STAR and/or PHENIX detector.

We present our results in three sections. In section 2 we discuss
prompt-$\gamma$ production per se. In section 3 we analyse
the additional information which is obtained by simultaneously
detecting the away-side-jet, and in section 4 we present the photon
asymmetry generated by an asymmetry in the production of $\pi^0$'s
subsequently decaying into photons.

\section{Prompt-$\gamma$-Production\label{pgp}}
The goal of the measurement of prompt-$\gamma$-production at RHIC is
the determination of the gluon polarisation $\Delta g$. For this it has
to be clarified that the signal is indeed proportional to $\Delta g$
and can be clearly separated from the background. For this purpose we
investigated the two leading processes
(i.e. first order in $\alpha_s$)
for prompt-$\gamma$-production,
namely the {\em Compton process} (see figure 1) and the
{\em annihilation process} (figure 2) and determined their
contribution at RHIC energies to the cross section for
different parton parametrisations.
Actually the hard matrix elements
for prompt-$\gamma$ production have been
calculated to NLO \cite{Vogel}, but {\sc PYTHIA} is set up in such a way
that the higher orders are effectively taken care of by the initial
and final state showering. As the NLO amplitudes do not show any
features qualitatively different from those of the LO ones this
procedure should be fine. Note however, that for heavy-quark
production the spin-effects could change substantially between LO and
NLO \cite{Kar}, which would require a more carefull treatment.\\
Furthermore we had a closer look at
the main contributions to the background and analysed the procedures
proposed in \cite{RSC} to discriminate it.
\subsection{The Compton process vs.\ the annihilation process}
The hadronic cross section for prompt-$\gamma$-production in
proton-proton collisions is given by
a convolution of the parton distributions and the partonic cross
section. In the spin averaged case
($\sigma=\frac{1}{2}\left(\sigma(\uparrow\downarrow)
+\sigma(\uparrow\uparrow)\right)$, where $\uparrow\downarrow$
($\uparrow\uparrow$) denotes antiparallel (parallel) spins of the two
protons) it reads
\begin{eqnarray}
&&E_\gamma\frac{{\rm d}\sigma_{pp\rightarrow \gamma
X}}{{\rm d}^3p_\gamma}(s,x_F,p_\perp)\label{gammaunpolall}\\
&=&
\sum_{ab}\int \ {\rm d} x_a  \ {\rm d} x_b \ P_a(x_a,Q^2) P_b(x_b,Q^2)
E_\gamma\frac{{\rm d}\hat{\sigma}_{ab\rightarrow\gamma
X}}{{\rm d}^3p_\gamma}(\hat{s},x_F,p_\perp),\nonumber
\end{eqnarray}
while for the spin difference
($\Delta\sigma=\frac{1}{2}\left(\sigma(\uparrow\downarrow)
-\sigma(\uparrow\uparrow)\right)$) it is given by
\begin{eqnarray}
&&E_\gamma\frac{{\rm d}\Delta\sigma_{pp\rightarrow \gamma
X}}{{\rm d}^3p_\gamma}(s,x_F,p_\perp)\label{gammapolall}\\
&=&
\sum_{ab}\int \ {\rm d} x_a  \ {\rm d} x_b \ \Delta P_a(x_a,Q^2) \Delta
P_b(x_b,Q^2)
E_\gamma\frac{{\rm d}\Delta\hat{\sigma}_{ab\rightarrow\gamma
X}}{{\rm d}^3p_\gamma}(\hat{s},x_F,p_\perp).\nonumber
\end{eqnarray}
Here the sum is over all partonic subprocesses which contributes to the
reaction $pp\rightarrow \gamma X$. $P_a$ and $P_b$ denotes the
unpolarised parton distributions of quarks and gluons, $\Delta P_a$
and $\Delta P_b$ the polarised ones. The latter are the difference
between partons of the same helicity as the hadron and those of
opposite helicity, the former are the sum
of the two helicities. The partonic cross sections
in the helicity averaged case is defined as
$\hat{\sigma}=\frac{1}{2}\left(\hat{\sigma}_{++}+\hat{\sigma}_{+-}\right)$
where the indices $+,-$ signifies the helicities of the incoming
partons. For $\Delta\hat{\sigma}$ the plus sign between the two terms
in the formula above
has to be replaced by a minus sign. $x_a$ and $x_b$ are the Bj{\o}rken
$x$-variables of the incoming partons, $p_\perp$ is the transverse
momentum of the outgoing photon in the $pp$-CMS. The longitudinal
momentum fraction of the $\gamma$ is defined as $x_F=2p^z_\gamma/\sqrt{s}$.
The polarised partonic cross section is given by
\begin{eqnarray}
E_\gamma\frac{{\rm d}\Delta\hat{\sigma}_{ab\rightarrow\gamma
X}}{{\rm d}^3p_\gamma}&=&\alpha\alpha_s\frac{1}{\hat{s}}
\left|\Delta
\overline{M}_{ab\rightarrow\gamma
X}\right|^2\delta\left(\hat{s}+\hat{t}+\hat{u}\right).
\end{eqnarray}
To obtain the unpolarised cross section one has simply to replace the
polarised matrix element
$\Delta\overline{M}_{ab\rightarrow\gamma X}$
by the unpolarised
$\overline{M}_{ab\rightarrow\gamma X}$.
Finally the partonic Mandelstam variables are related to the usual
hadronic ones by: $\hat{s}=x_ax_bs$, $\hat{t}=x_at$, and $\hat{u}=x_bu$.

In leading order perturbation theory in $\alpha\alpha_s$ only the
Compton process $qg\rightarrow \gamma q$ and the annihilation process
$q\bar{q}\rightarrow \gamma g$ contribute to the
prompt-$\gamma$-production. Their polarised and unpolarised matrix
elements are summarised in table 1.
The contribution of the two processes to the hadronic cross section
corresponding to (\ref{gammaunpolall}) resp.\ (\ref{gammapolall})
depends strongly on the parton distributions. In {\em unpolarised}
$pp$-collisions -- in contrary to $p\bar{p}$-collisions -- the Compton
process clearly dominates over the annihilation process, because the
gluon density in protons is much higher than the antiquark density
($g\gg\bar{q}$). In the polarised case, however, the relative
importance of the two processes depends crucially on the relative
size of
the polarised gluon distribution $\Delta g$ and the
polarised sea distribution $\Delta \bar{q}$. Both of them are
presently completely unknown. By the time the RHIC-Spin-Collaboration
(RSC)
could possibly start to take data, $\Delta \bar q(x)$ should,
however, be known quite accurately from semi-inclusive lepton-nucleon
scattering experiments by the HERMES collaboration (HERA)
\cite{HERMES}.
This should allow to avoid ambiguities in the interpretation of
potential RSC data.\\
For our simulation we used two parametrisations
for parton densities with large gluon
polarisation by
Altarelli\&Stirling \cite{altsti} and by Ross\&Roberts (set D)
\cite{rosrob}
and one parametrisation with a small gluon polarisation by
Ross\&Roberts (set A). \\
For large gluon polarisation the Compton process is the by far dominant one
and one can safely neglect the contribution of the annihilation
process in (\ref{gammapolall}). In this case the
prompt-$\gamma$-production becomes proportional to $\Delta g$ and is
such a clean probe for the gluon polarisation:
\begin{eqnarray}
&&E_\gamma\frac{{\rm d}\Delta\sigma_{pp\rightarrow \gamma
X}}{{\rm d}^3p_\gamma}(s,x_F,p_\perp)\label{gammapol}\\
&\approx&
\sum_{q}\int {\rm d} x_a  \ {\rm d} x_b \left(\Delta q(x_a,Q^2) \Delta
g(x_b,Q^2)
E_\gamma\frac{{\rm d}\Delta\hat{\sigma}_{qg\rightarrow\gamma
q}}{{\rm d}^3p_\gamma}(\hat{s},x_F,p_\perp) +(x_a\leftrightarrow
x_b)\right).\nonumber
\end{eqnarray}
However, in a scenario with a large sea contribution to the spin of
the proton and a gluon polarisation only due to Altarelli-Parisi
evolution, as described by the parametrisation Ross\&Roberts set A
\cite{rosrob},
the annihilation process becomes the major contribution.

We investigated these different scenarios with the Monte-Carlo program
{\sc Sphinx}, which can be used to simulate longitudinal
polarised $pp$-scattering. We generated $10^7$ events for both spin
combinations of the protons at the RHIC energy $\sqrt{s}=200\ {\rm
GeV}$. For the unpolarised parton distributions we haven chosen the
parametrisation of Gl\"uck, Reya, and Vogt \cite{grv}, while for the
polarised distributions we used the parametrisations mentioned above,
namely Altarelli\&Stirling and Ross\&Roberts set A and set D. In {\sc
Sphinx} matrix elements are implemented in leading order only. However,
due to the initial and final state shower algorithm some features of
higher order effects are incorporated as well \cite{pyman}. Also the
polarisation effects are traced in the initial state shower.
For the simulations the polarised initial state shower and the final
state shower were switched on. To avoid infrared divergences
the hard interaction cross section must be supplemented by a lower
cut off for the transverse momentum $p_\perp$. We chose
$p_\perp\ge4\ {\rm GeV}$.

The results of these simulations are shown in figures 3 to 21.
In
figure \ref{fig3} the Lorentz-invariant cross section for
prompt-$\gamma$-production as a function of $p_\perp$ at $x_F\approx 0$
is displayed for the Compton process (upper plot) and the annihilation
process (lower plot).
$x_F$ is here the longitudinal momentum fraction of the photon
defined by $x_F=2p_z^{\gamma}/\sqrt{s}$.
In both cases the spin averaged cross section
(squares) and the cross section for the spin difference (triangles)
are shown. For the latter the parametrisation of Altarelli\&Stirling
has been used. For the annihilation process $-\Delta\sigma$ is
plotted, because $\Delta\sigma$ is negative, meaning that the
cross-section for antiparallel spins is smaller than for parallel
spin. This can be seen from table 1, keeping in mind that
$\hat t =x_1x_3 t =-\hat s(1-\cos \theta)/2$ is negative.
This negative
polarised partonic cross section is than multiplied by  the positve
polarised quark and antiquark distributions in (\ref{gammapolall}).
The error bars reflect the MC error.
A typical RHIC run has
320 pb$^{-1}$, such that the $10^7$ events we generated for each spin
combination
(with $p_{\perp}\ge 4$ GeV)
corresponds to an integrated cross section of $3
\cdot 10^{-5}$ mb for the spin averaged case respectively to a
differential cross-section of roughly $3\cdot 10^{-5}~ {\rm mb}/
(4\pi\cdot 4 {\rm
 ~GeV})=6 \cdot 10^{-7}$ mb/GeV in the 4 GeV bin. This implies that the MC
error is roughly comparable to the expected experimental error for
the prompt-$\gamma$'s and substantially larger than the
anticipated experimental errors for gammas from $\pi^0$ decays.

At this point we want to state as clearly as possible that we do not
attribute special significance to any of the used parametrisations.
In fact virtually nothing is known about $\Delta g(x,Q^2)$, except
the trivial fact that its absolute magnitude is limited by
$g(x,Q^2)$. It could e.g. very well be that $\Delta g(x,Q^2)$ changes
sign for some $x$ value and this was actually advocated to get a good
fit to the data in specific models. It is just because nothing is known
about $\Delta g(x,Q^2)$ that it is so important to measure (and RHIC
seems to be the only facility able to do so). In this situation all
our simulations are meant as illustrations. We used two similar
models (Altarelli\&Stirling and Ross\&Roberts D) with optimistically
large gluon-distributions, because it is easy to estimate the results
for smaller $\Delta g$ by just scaling them down and for large
$\Delta g$ the MC simulation requires less statistics. At the same
time the difference between the predictions of these two models gives
a feeling for the sensitivity of experimental signals to details
of the distribution functions.

Comparing the two processes one can clearly see that in the spin
averaged case the Compton process leads to a cross section by an
order of magnitude larger than the annihilation process. This
domination is even more pronounced in the polarised case, where one
and a half orders of magnitude are between both processes. However,
the latter fact is, as said above, due to the choice of a parametrisation
with a large $\Delta g$ and a small $\Delta \bar{q}$.

In figure \ref{fig4} we examine the polarised case in more detail by using
different sets of polarised parton distributions.
In the upper part the cross section of the spin difference for the
Compton process for three different sets of parametrisations
is shown. The parametrisations of Altarelli\&Stirling (triangles) and
Ross\&Roberts set D (squares) have a similar, large $\Delta g$, but
the $x$-dependence $\Delta g(x)$ is different, which leads to much
bigger cross section (logarithmic scale!) in the latter case in
comparison with the former. This means that it is possible to
distinguish between both parametrisations in the experiment. The third
parametrisation Ross\&Roberts set A (circles) has a gluon polarisation
only due to Altarelli-Parisi evolution and leads to no measurable
signal.

In the lower part the annihilation process is shown. Here the results
for Altarelli\&Stirling (triangles), as a representative of the large
polarised
gluons and small polarised sea scenario, and Ross\&Roberts set A
(circles), as a
representative of the opposite situation of nearly unpolarised gluons
and large, negative polarised sea, are given. As seen above,
Altarelli\&Stirling leads to a small, negative signal, whereas now
Ross\&Roberts set A produce a relatively large, {\em positive} signal,
which exceeds the corresponding result for the Compton process by far.
On the other hand it is still by roughly a factor of five smaller than
the result of Compton process in case of a large gluon polarisation.

In figure \ref{fig5} the resulting asymmetries $\Delta\sigma/\sigma$ are
shown. For the Compton process in the upper part the asymmetry grows
with $p_\perp$ for Altarelli\&Stirling and Ross\&Roberts set D and
reaches values around 20\% resp.\ 30\%. For Ross\&Roberts set A it is
consistent with zero.
For the annihilation process in the lower part Altarelli\&Stirling
leads to small, negative, Ross\&Roberts set A to a large, positive
asymmetry.

The main conclusions one can draw at this point are that in the case
of a large gluon polarisation the annihilation process is a small
correction, which is  calculable using the data from electron-scattering,
such
that prompt-$\gamma$-production is a clean probe of
$\Delta g$.
If the absolute value of $\Delta g(x)$ would be a factor ten
smaller than assumed by e.g. Altarelli\&Stirling, which would be the case
if the total spin carried by gluons were less than half a unit of
$\hbar$,  then annihilation and Compton graph would contribute at the
same level. It would still be possible to extract $\Delta g(x,Q^2)$,
but only by a combined fit to the data from various spin-experiments.
If $\Delta g$ would be substantially smaller than 0.5 $\hbar$ the
determination of $\Delta g(x,Q^2)$ were probably very difficult
(unless rather large positive and negative parts of $\Delta g(x)$
 canceled to give
a small $\Delta g$).
However, deriving such a low bound for $\Delta g(x)$ would be very
interesting as one would not expect it to be that small. The
crucial remaining questions are whether the anticipated statistics is
sufficent to actually determine $\Delta g(x)$ and whether there are
background effects
which could blur the simple picture. These questions will be
addressed next.

To examine the precission of
prompt-$\gamma$ measurements at RHIC, we transformed
these cross sections in counting rates at RHIC and determined the
statistical
errors. To obtain the total rates one has to multiply the cross
section with the integrated luminosity for which we have taken the design
value from \cite{RSC}: $\int{\cal L}{\rm
d}\tau=3.2\times10^{38}\ {\rm cm}^{-2}$, which corresponds to a
luminosity of ${\cal L}=8\times 10^{31}\ {\rm cm}^{-2}\ {\rm s}^{-1}$
and an effective run time of $\tau=4\times 10^6 \ {\rm s}$, which
means 100 days with 50\% efficiency. At RHIC the beams are only
partially polarised: $P_{\rm beam}=0.7$, whereas {\sc Sphinx}
simulates fully polarised events. Therefore one has to combine the
MC-rates to the experimental rates as follows ($P=P_{\rm beam}\times
P_{\rm beam}$):
\begin{eqnarray}
N_{\rm exp.}^{\uparrow\downarrow}&=&
\frac{1+P}{4}N_{\rm MC}^{\uparrow\downarrow}
+\frac{1-P}{4}N_{\rm MC}^{\uparrow\uparrow}\\
N_{\rm exp.}^{\uparrow\uparrow}&=&
\frac{1+P}{4}N_{\rm MC}^{\uparrow\uparrow}
+\frac{1-P}{4}N_{\rm MC}^{\uparrow\downarrow}.
\end{eqnarray}
For the asymmetry follows:
\begin{eqnarray}
A_{\rm exp.}&\equiv&
\frac{N_{\rm exp.}^{\uparrow\downarrow}-
N_{\rm exp.}^{\uparrow\downarrow}}{N_{\rm exp.}^{\uparrow\downarrow}
+N_{\rm exp.}^{\uparrow\downarrow}}\\
&=&P\frac{N_{\rm MC}^{\uparrow\downarrow}-
N_{\rm MC}^{\uparrow\downarrow}}{N_{\rm MC}^{\uparrow\downarrow}
+N_{\rm MC}^{\uparrow\downarrow}}\\
&\equiv&PA_{\rm MC},\label{asymcexp}
\end{eqnarray}
with the error:
\begin{eqnarray}
\delta A_{\rm exp.}&=&\frac{2
\sqrt{N_{\rm exp.}^{\uparrow\downarrow}N_{\rm exp.}^{\uparrow\uparrow}}}{
N_{\rm exp.}^{\uparrow\downarrow}+N_{\rm exp.}^{\uparrow\uparrow}}
\frac{1}{\sqrt{N_{\rm exp.}^{\uparrow\downarrow}+
N_{\rm exp.}^{\uparrow\uparrow}}}\\
&\approx&
\frac{1}{\sqrt{N_{\rm exp.}^{\uparrow\downarrow}+
N_{\rm exp.}^{\uparrow\uparrow}}}.\label{asyerr}
\end{eqnarray}
At RHIC several cuts on the events have to be applied. One cut, which
we have investigated further, is due to the finite coverage in the
plane of the
pseudorapidity $\eta$ and the azimuth $\phi$ of the detectors STAR and
PHENIX \cite{RSC}.
STAR has a full coverage in the azimuth: $\Delta \phi=2\pi$ and a
coverage in the pseudorapidity of $|\eta|<1$ without end caps and of
$|\eta|<2$ in the extended version with end caps. PHENIX covers
only the half azimuth $\Delta \phi=\pi$ and has the same
$\eta$-coverage as STAR. To examine how severe these cuts are we
determined the rates for prompt-$\gamma$-production assuming a full
azimuthal coverage\footnote{This means that we are simulating the
situation for STAR. To obtain the corresponding results for PHENIX one
has to divide the rates by 2.} and applying the cuts
$|\eta^\gamma|<1$, $|\eta^\gamma|<2$, and no $\eta^\gamma$-cut.
The results are given in the figures \ref{fig6},\ref{fig7}.
The error bars reflect
the statistical errors at RHIC.

In figure \ref{fig6} the spin averaged rates (upper left plot), the spin
difference rates (upper right plot), and the asymmetry (lower plot)
are displayed for the three $\eta^\gamma$-cuts. The parametrisation of
Altarelli\&Stirling has been used.
One can clearly see that the cut $|\eta^\gamma|<2$ is
not very severe, especially for high $p_\perp$, whereas the cut
$|\eta^\gamma|<1$ reduces the rates roughly by a factor of two. In
addition also the asymmetry decreases in this case slightly.
However, even with the stronger cut $|\eta^\gamma|<1$ similar
parametrisations as Altarelli\&Stirling and Ross\&Roberts set D can be
distinguished within the experimental errors at RHIC, as can be seen in
figure \ref{fig7}. Here the rates for the spin difference and the
asymmetries for the different parametrisations in dependence of the
applied cuts are compared. Hence at this point the extended version of
detectors with end caps is not absolutely needed, but, as shown later,
this extension will be crucial when measuring the prompt-$\gamma$ and
the away-side jet.
\subsection{Background considerations}
High-$p_\perp$ $\gamma$'s are not only produced in the direct
processes discussed in the subsection above, but at a far larger
rate due to bremsstrahlung and in particular in meson decays. This
background has to be separated from the direct photons very accurately
in order to do not contaminate the signal substantially. In this publication
we
concentrate on the background produced in pion and $\eta$ decays
$\pi^0\rightarrow2\gamma$ resp.\
$\eta\rightarrow2\gamma$ which is the major contribution and analyse
the ability to remove it with STAR and PHENIX.

The mesons which give rise to the background are produced in all QCD
parton processes, among them the QCD-Compton process:
\begin{eqnarray*}
qg\rightarrow gq&\longrightarrow& \dots \pi^0\rightarrow \dots \gamma\gamma,\\
qg\rightarrow gq&\longrightarrow& \dots \eta\rightarrow \dots\gamma\gamma
\end{eqnarray*}
is the most important, followed by $qq\rightarrow qq$ and
$gg\rightarrow gg$. Beside the QCD-Compton process we had a closer
look at the $gg$-scattering process, because this process is
especially interesting with regard to the gluon polarisation. Here
$\Delta g$ enters twice in the cross section, and hence this process
is extremly sensitive on changes in $\Delta g$.
For our studies we simulated $10^7$ events for each polarisation
combination of the protons for both, the QCD-Compton process
$qg\rightarrow gq$ and the $gg$-scattering process $gg\rightarrow gg$.

Comparing the yields of the QCD-Compton process (full symbols) with the
prompt-$\gamma$ Compton process (open symbols) in figure \ref{fig8} it is
obvious that the background is very important in the case of a large
gluon polarisation (Altarelli\&Stirling left, Ross\&Roberts right).
In the spin averaged case (upper plots) it is much larger
than the signal, but it decreases also faster with $p_\perp$. In
addition, the background shows also an asymmetry and the resulting
rates for the spin-difference (lower plots) are higher than the true
signal up to a
transverse momentum of $p_\perp\approx 15 \ {\rm GeV}$.
Because the QCD-Compton process is just one contribution, although the most
important, to the background, it is obvious that the
photons from neutral meson decay have to be separated very accurately from the
direct ones in order not to contaminate the true signal. Doing so has
the additional advantage that the asymmetry in the photons from e.g. pion
decay can be used as an excellent signal for $\Delta g(x)$ (see
section 4).

To investigate the background further let us consider how it is composed.
Figure \ref{fig9} and \ref{fig10} show the compositon for the QCD-Compton
process and the
$gg$-scattering process. In both cases the the pion decay
$\pi^0\rightarrow 2\gamma$ give rise to the main contribution of
80\%-90\% (logarithmic scale!) over the whole $p_\perp$-region for
both, the spinaveraged
case (upper plots) and the spin difference (lower plots). The next
important contribution is the $\eta$-decay $\eta\rightarrow 2\gamma$
with 10\%-20\%.

Comparing figure 9 and figure 10 one realizes that the rate of
the QCD-Compton process is several times larger than that of the
$gg$-scattering process in the spin averaged case (upper plots).
For the spin-difference they are comparable at low $p_\perp$ even with
the parametrisation of Altarelli\&Stirling with its large
gluon polarisation. Due to the much faster decrease with $p_\perp$ of
the latter the QCD-Compton process becomes also in the polarised case
the dominant one at high-$p_\perp$. This dominance will be more
pronounced for parametrisations with a smaller $\Delta g$.

 From the discussion above it follows that the mesons
from which those background photons stem have to be reconstructed,
in order that they can be
separated from the true signal. There are two main possibilities that
the mesons escape their reconstruction and thus `fake' photons remain. The
two possible sources of `fake' $\gamma$'s are the following.
First, {\em asymmetric decay}, i.\ e.\ one $\gamma$ is
inside the detector, the other outside, such that they cannot be
combined.
Second, {\em merged $\gamma$'s}, i.\ e.\ the two $\gamma$'s are too
narrow and cannot be resolved by
the detector. Investigations have been done only for the latter case,
because this is more general, whereas the former is dependent on
the precise detector geometry and
requires a specialized  detector simulation.

In the following we determine the rate of `fake' $\gamma$'s from
pion decay in dependence of the spatial detector resolution.
The minimal opening angle of a $\gamma$-pair in the
rest frame of the pion is given by:
\begin{eqnarray}
\chi_{\rm min}&=&2\frac{m_\pi}{E_\pi}.\label{chimin}
\end{eqnarray}
The following resolutions of the detector are considered
\begin{eqnarray}
\chi^{\rm res}&>&0,005\ {\rm rad}\nonumber\\
\chi^{\rm res}&>&0,01\ {\rm rad}\nonumber\\
\chi^{\rm res}&>&0,02\ {\rm rad},
\end{eqnarray}
For the PHENIX detector the planned design would result
in $\chi^{\rm res}>0,01\ {\rm rad}$.
We define the fake-$\gamma$-rate $R$ as the fraction of the number of
unresolved pions and the total number of pions:
\begin{eqnarray}
R&=&\frac{N_\pi^{\rm unres}}{N_\pi}
\end{eqnarray}
The MC fake-$\gamma$-rate $R_{\rm MC}$ has to be transformed to the
experimental $R_{\rm exp}$ as follows:
\begin{eqnarray}
R_{\rm exp}^{\uparrow\downarrow(\uparrow\uparrow)}
&=&\frac{R_{\rm MC}^{\uparrow\downarrow(\uparrow\uparrow)}
+\rho R_{\rm MC}^{\uparrow\uparrow(\uparrow\downarrow)}}{1+\rho},
\end{eqnarray}
with
\begin{eqnarray}
\rho&\equiv&\frac{1-P}{1+P}\times \frac{1-A_{\rm MC}}{1+A_{\rm MC}}.
\end{eqnarray}
Then the number of real direct photons is given by:
\begin{eqnarray}
N_\gamma&=&N_{\rm tot}-R N_\pi.
\end{eqnarray}
The total number of fake-$\gamma$'s in dependence of their opening angle
and the fake-$\gamma$-rates for
the different resolutions considered of the detector are
displayed in figure \ref{fig11}. These plots show that the PHENIX
resolution of $\chi^{\rm res}>0,01$ is sufficient to keep the
fake-$\gamma$-rate
below 10\% up to a transverse momentum of $p_\perp\approx 20\ {\rm GeV}$.
Due to the steep slope of the $\gamma$-rates around $\chi=0,01$ in the
upper plots the resolution $\chi^{\rm res}>0,02$ is absolutely
unsuited, whereas
the better resolution $\chi^{\rm res}>0,005$ is excellent. Further studies are
needed to determine the optimal compromise between costs and
fake-$\gamma$-rates for the
detector.
In addition there is
a slight spin dependence of the fake-$\gamma$-rates
$R^{\uparrow\uparrow}<R^{\uparrow\downarrow}$ such that unless
$\chi^{\rm res}=0,005$ a carefull simulation is needed to
determine the resulting corrections and thus keep the
systematic uncertainties small.
\section{Prompt-$\gamma$ and Away-Side Jet}
Additional information can be extracted from prompt-$\gamma$
measurements if they are observed in coincidence with jets, which
could also help to reduce unwanted background. The
resulting information could e.g. be used
to determine the $x$-dependence of the gluon polarisation $\Delta
g(x)$ which is not sufficiently determined by the prompt-$\gamma$
signal alone. In practice, however, one will determine all the
polarised distribution functions by simultaneous fits to all data. In
the context of such fits prompt-$\gamma$'s plus jet data could he of
great importance for $\Delta g(x,Q^2)$.  \\
Let us describe first how we handled the jet-reconstruction.
Our Monte-Carlo code just like {\sc PYTHIA} uses the jet routines of
{\sc Jetset}. Jet reconstruction was thus done by the {\sc
Jetset}-subroutine
{\tt LUCELL} \cite{Sjo92}.  This routine is also used to analyse
unpolarised proton-proton collisions and defines jets in the
two-dimensional $(\eta-\phi)$- plane, $\eta$ being the rapidity and
$\phi$ the angle around the $z$-axes. For our calculations we used 25
$\eta$-bins and 24 $\phi$-bins with various bounds for $|\eta |$. The
jet defining algorithm works as follows. First all transverse
energies  $E_{\perp}=\sqrt{p_{\perp}^2+m^2}$ in a bin are summed. If
this sum exceeds a certain value ($E_{\perp}^{\rm cell}$),
which we chose to be 1.5 GeV, than it is treated as a jet candidate.
Starting from the cell with the highest transverse energy all cells
in a smaller `distance'  than
$R=\sqrt{(\Delta\eta)^2+(\Delta\phi)^2}$
(which we chose as 0.7) are combined to a `cluster' and if
the total transverse energy in this cluster exceeds
$E_{\top}^{\rm cluster}$ which we chose to be 3.5 GeV, the
contained
particles are accepted as a jet.

Finally we use an additional constraint on $|\eta|$ implied by the
present design of the STAR detector. This detector is planned to be
built with so-called end-caps, giving them a wider $\eta$-range.
With these end-caps it covers the range
$\left|\eta^\gamma\right|\leq1$, $\left|\eta^{\rm Jet}\right|\leq0.3$
while without them only the range
$\left|\eta^\gamma\right|\leq2$, $\left|\eta^{\rm Jet}\right|\leq1.3$.
is accessible.

We start by showing that the larger $\eta$ range is really needed to
do allow for a sensible jet analyses.
Figure 12 shows a rapidity distribution of the jets generated by our
code and the consequences of the rapidity cuts. In each case the
area of one of the small rectangels measures the logarithm of the
number of events. Obviously the rapidity cuts without end-caps
are too restrictive while adding the end-caps allows to cover nearly
all of the interesting rapidity-range.\\

We are interested in the Compton-process of figure 1. Thus we require
that the jet and the photon are detected with a relative angle above
90 degrees in the partonic center of momentum system. In the
following,
jets fullfilling this criterium
are called away-side jets. In the hadronic center of momentum system,
which for RHIC coincides with the laboratory system this condition
looks rather complicated due to the Lorentz-boost
\begin{equation}
\cos \chi= {4x_1x_2\cos(\phi^{\gamma}-\phi^{\rm Jet})+
(x_2e^{\eta^{\gamma}}-e^{-\eta^{\gamma}})
(x_2e^{\eta^{\rm Jet}}-e^{-\eta^{\rm Jet}}) \over
(x_2e^{\eta^{\gamma}}+e^{-\eta^{\gamma}})
(x_2e^{\eta^{\rm Jet}}+e^{-\eta^{\rm Jet}})} <0
\end{equation}
but it turned out that for the jets we generated this criterium is
actually equivalent to the much simpler one
\begin{equation}
\cos(\phi^{\gamma}-\phi^{\rm Jet})<0 ~~~~~~~.
\end{equation}
The consequences of this criterium are shown in figure 13. In this
figure we show histograms of the generated jets as a function of the
photon and jet rapidity. If the `away-side-jet' criterium is not used
we get the results on the left side for the spin averaged and spin
difference rates. Obviously the photons are in general very strongly
correlated with the jet axes. If the criterium is applied to single
out the hard Compton processes the distributions on the right side
are obtained. Obviously these are only a small fraction of the total
events and the rapidities are substantially different. The weak
remaining correlation is a consequence of the Lorentz-boost. In the
partonic center of momentum system the rapidities are
anti-correlated.

The cross sections for photon-plus-jet events are related to the
distribution functions according to
\begin{eqnarray}
\frac{d^3\sigma(pp\rightarrow \gamma +{\rm Jet}+X)}{dp_\perp^2
d\eta^\gamma d\eta^{\rm Jet}} \sim \sum_q e_q^2 q(x_a)
g(x_b)\frac{d\sigma(qg\rightarrow\gamma q)}{d\hat{t}}
 ~~~~~~~\\
\frac{d^3\Delta\sigma(pp\rightarrow \gamma +{\rm Jet}+X)}{dp_\perp^2
d\eta^\gamma d\eta^{\rm Jet}} \sim \sum_q e_q^2 \Delta q(x_a)
\Delta g(x_b)\frac{d\Delta\sigma(qg\rightarrow\gamma q)}{d\hat{t}}
\nonumber
\end{eqnarray}
 From this we can calculate the differential
$\gamma$-jet cross sections as a function of $\eta^{\gamma}$ and
$\eta^{\rm Jet}$. Figure 14 and 15 show the results for spin average
and spin difference in the $\eta^{\gamma}$=0,
$\eta^{\rm Jet}$=0 bin for our two parametrisations with large
$\Delta g$.

Photon-jet experiments give actually more information than contained
in figure 14 and 15. It is possible to reconstruct the $x$-values
from the measured pseudorapidities and transverse momenta. In the
ideal case, neglecting all initial and final state interaction this
connection is simply given by
\begin{eqnarray}
p_1&=&x_1\left(P,0,0,P\right)\nonumber\\
p_2&=&x_2\left(P,0,0,-P\right)\nonumber\\
p^\gamma&=&p_\perp\left(\cosh\eta^\gamma, \cos\phi^\gamma, \sin\phi^\gamma,
\sinh\eta^\gamma\right)\nonumber\\
p^{\rm Jet}&=&p_\perp\left(\cosh\eta^{\rm Jet}, \cos\phi^{\rm Jet},
\sin\phi^{\rm Jet}, \sinh\eta^{\rm Jet}\right).
\end{eqnarray}
where $p_1$ and $p_2$ are the four-momenta of the incomming partons.
Energy-momentum conservation implies than (neglecting all masses)
\begin{eqnarray}
x_1&\simeq&\frac{2p_\perp}{\sqrt{s}}\left(\frac{e^{\eta^\gamma}+e^{\eta^{\rm
Jet}}}{2}\right)\nonumber\\
x_2&\simeq&\frac{2p_\perp}{\sqrt{s}}\left(\frac{e^{-\eta^\gamma}+e^{-\eta^{\rm
Jet}}}{2}\right). \label{xrek}
\end{eqnarray}
Finally to decide which of these x-values belongs to the gluon and
which to the quark the following procedure was suggested: Define
$x_a={\rm min}(x_1,x_2)$ and
$x_b={\rm max}(x_1,x_2)$ and require $x_b\ge 0.2$. As the gluon
distribution is already small at such $x$-values one can expect that
$x_b$ is the quark momentum fraction.

In proceeding like this a number of rather severe asumptions were
made such that it was rather unclear how good it would work. {\sc
SPHINX}
gives us the possibility to check it explicitely. Figure 16 to 19 show
the results. Figure 16 shows just histograms of the generated quark
and gluon momentum fractions. Figure 17 shows how these are
correlated with $x_a$, $x_b$, $x^{\rm exp}_g$, $x^{\rm exp}_q$.
Again the area of the rectangles is proportional to the logarithm of
the rate, such that the correlations are much stronger than they
look.
The
figures on the left side show how the $x$ values of the distribution
functions $x^{\rm MC}$ are correlated to those generated by the
complete {\sc SPHINX} algorithm, i.e. the difference between
$x^{\rm exp}_g$ and $x_g$ respectively $x^{\rm exp}_q$ and $x_q$ is
entirely due to the intrinsic transverse momentum and to effects of
the initial and final state showering. The identification of the jets
is taken from the Monte-Carlo, such that there are no
misidentifications.
In an ideal experiment $x^{\rm exp}_g$ and $x^{\rm exp}_q$  are the best
measurable approximations for $x_g$ and $x_q$.  The right side shows how
good the simple procedure just
described is able to reconstruct the $x$-values, still without the
$x_b$-cut. Obviously the quark momentum fraction can be reconstructed
quite reasonably, while the reconstruction of $x_g$ is problematic.
This is why a $x_b$ cut is needed. Introducing it the correlation
becomes much better, as shown in more detail in figure 18 and 19 for
$g(x)$ and $\Delta g(x)$. In these figures the upper graphs shows the
histogram of the $x$ values actually chosen by the Monte Carlo code.
The lower ones show the distribution of $x$-values reconstructed with
the described procedure. The agreement is very good for $x>0.04$ and
gets rapidly bad if one goes to smaller $x$-values. The
problems are far less pronounced for the
polarised case because the chosen function for $\Delta g(x)$ is
comparably small at small $x$. For a situation where $\Delta g(x)$
would be concentrated at extremely small $x$ it could not be deduced
from the $\gamma$-jet signal. The latter is however true for all
polarised experiments. Such a very soft polarised gluon distribution
would also not be detected in deep inelastic lepton-nucleon
scattering such that it could not help to explain the
observed
data. Its only effect could be to screw up the extrapolation to
small-$x$ needed to derive experimental values for the sum rules.
If the observed data are interpreted as giving evidence for an
anomalous gluon contribution (however one is trying to define it)
than $\Delta g(x)$ cannot be too soft and thus should be seen by
RHIC.

With all these caveats one should note, however, that our simulation
shows, that an actual prompt-$\gamma$-plus-jet experiment would be
able to distinguish even between the two rather similar gluon
distributions we used (see figure 19, the bottom plots).

We conclude this paragraph by stating that prompt-$\gamma$-jet
coincidences will give interesting data but that their analyses will
be highly non-trivial requirering extensive numerical simulations.
A cut like $x_b\geq 0,2$ is necessary. While a more restrictive cut
leads to a better reconstruction it also
worsens statistic. We hope that {\sc SPHINX} will help to find the best
compromise.

\section{$\pi^0$-Production\label{pip}}
We discussed already in section 2 that the $\pi^0$ decaying into two
photons have to be reconstructed in order to extract the
direct-$\gamma$ signal. This should be possible with high efficency.
However, this opens also the possibility to use the $\pi^0$ asymmetry
as
an independent measurement of the
gluon polarisation as well. Their advantage in comparison with the
prompt-$\gamma$'s
is the larger cross section and hence the smaller statistical error.
On the other hand the pions are a less direct probe for $\Delta g$
because several processes contribute to their rate, such that the
total observed asymmetry depends in a rather involved manner on the
gluon polarisation. For example the most
important contribution, the QCD-Compton process is proportional to
$\Delta g$, but the next important one $qq\rightarrow qq$ is
independent of $\Delta g$, whereas in $gg\rightarrow gg$ it enters
twice.
Again a detailed Monte-Carlo simulation should allow to relate the
data to the distribution functions.

In figure \ref{fig20} the contributions of the QCD-Compton process and the
$gg$-scattering process to the $\pi^0$-production are displayed.
In the spin averaged case the QCD-Compton process is the far more
important. Due to the large gluon polarisation in Altarelli\&Stirling,
which enters in
the $gg$-scattering process twice, this process gives the major
contribution at low $p_\perp$. However, because it decreases much
faster with $p_\perp$, for $p_\perp>8\ {\rm GeV}$ the QCD-Compton
process becomes the dominant one in the polarised case too, even for
large gluon polarisation. Nevertheless, the $gg$-scattering process
leads to an asymmetry much larger than the QCD-Compton process.
Here, the MC-statistic gives reliable predictions only up to
$p_\perp<15\ {\rm GeV}$.
The different $p_\perp$-dependence of the individual processes could
be used to disentangle them.

Figure 21 shows the yield of pions due to the QCD-Compton process
in the spin averaged case (upper
plot), for the spin-difference (middle) and the resulting asymmetry
(lower plot) as
a function of $p_\perp$ for parametrisations with a large gluon
polarisation.
While figure 20 showed results for $x_F=0$ figure 21 shows the rates
integrated over $x_F$.
In comparison with the prompt-$\gamma$-datas the two
parametrisation of Altarelli\&Stirling (squares) and Ross\&Roberts set
D (triangles) are much harder to distinguish on the base of these rates.
Although this is unfortunate in this case it shows that one is really
analysing an observable which tests different properties than
the prompt-$\gamma$-measurements.

\section{Summary}
We have analysed three observables, namely direct
prompt-$\gamma$'s, prompt-$\gamma$'s in coincidence with an away-side
jet, and $\gamma$'s from $\pi^0$-decay.
We found that each of them should allow to obtain significant
results in RHIC spin-physics experiments. These results should allow
to settle the problem of $\Delta g(x)$. However, each of these
signals has its problems: For direct photons the background rate in
rather high, for prompt-$\gamma$-jet coincidences the jet
reconstruction is non-trivial and has, just as for the $\pi^0$-decay,
to rely very heavily on Monte-Carlo simulations. Consequently the
existence of several independent codes
will be crucial to relate the observed data in a reliable way to the
basic physical quantities of interest. We hope that our code will
contribute to this endeavour.

\section{Acknowledgement}
We are very much indepted to T. Sj\"ostrand for his help in
understanding the details of {\sc PYTHIA} and for his encouragement. A.S.
thank DFG, BMFT, and the MPI Heidelberg for support. L.M. has been
supported by KBN grant 2-P302-143-06.

\pagestyle{empty}

\clearpage
\begin{table}
\begin{tabular}{|lrr|}
\hline
&&\\
Subprocess&$|\overline{M}|^2$&$|\Delta\overline{M}|^2$\\
&&\\
\hline
&&\\
$qg\rightarrow \gamma q$&
$-e_q^2\frac{1}{3}~(\hat{s}^2+\hat{t}^2)/(\hat{s}\hat{t})$&
$-e_q^2\frac{1}{3}~(\hat{s}^2-\hat{t}^2)/(\hat{s}\hat{t})$\\
&&\\
\hline
&&\\
$q\bar{q}\rightarrow \gamma g$&
$e_q^2\frac{8}{9}~(\hat{t}^2+\hat{u}^2)/(\hat{t}\hat{u})$&
$-e_q^2\frac{8}{9}~(\hat{t}^2+\hat{u}^2)/(\hat{t}\hat{u})$\\
&&\\
\hline
\end{tabular}
\label{table1}
\end{table}
\vskip 3 cm
{\bf Table 1:~~ The squarred matrix elements for the Compton- and
annihilation process}

\clearpage
\begin{figure}[t]
  \centering
  \makebox[14cm]{\epsfysize=4.cm
                  \epsffile{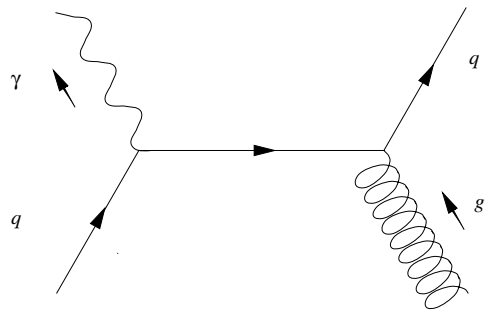}}
  \caption[Der Compton-Proze{\ss}]{{\small\sl The Compton graph
  \label{fig1}}}
\end{figure}

\clearpage
\begin{figure}[t]
  \centering
  \makebox[14cm]{\epsfysize=4.cm
                  \epsffile{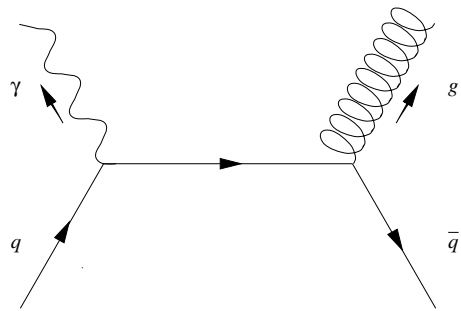}}
  \caption[Der Annihilationsproze{\ss}]{{\small\sl The annihilation
graph
  \label{fig2}}}
\end{figure}

\clearpage
\begin{figure}[htb]
  \centering
  \epsffile{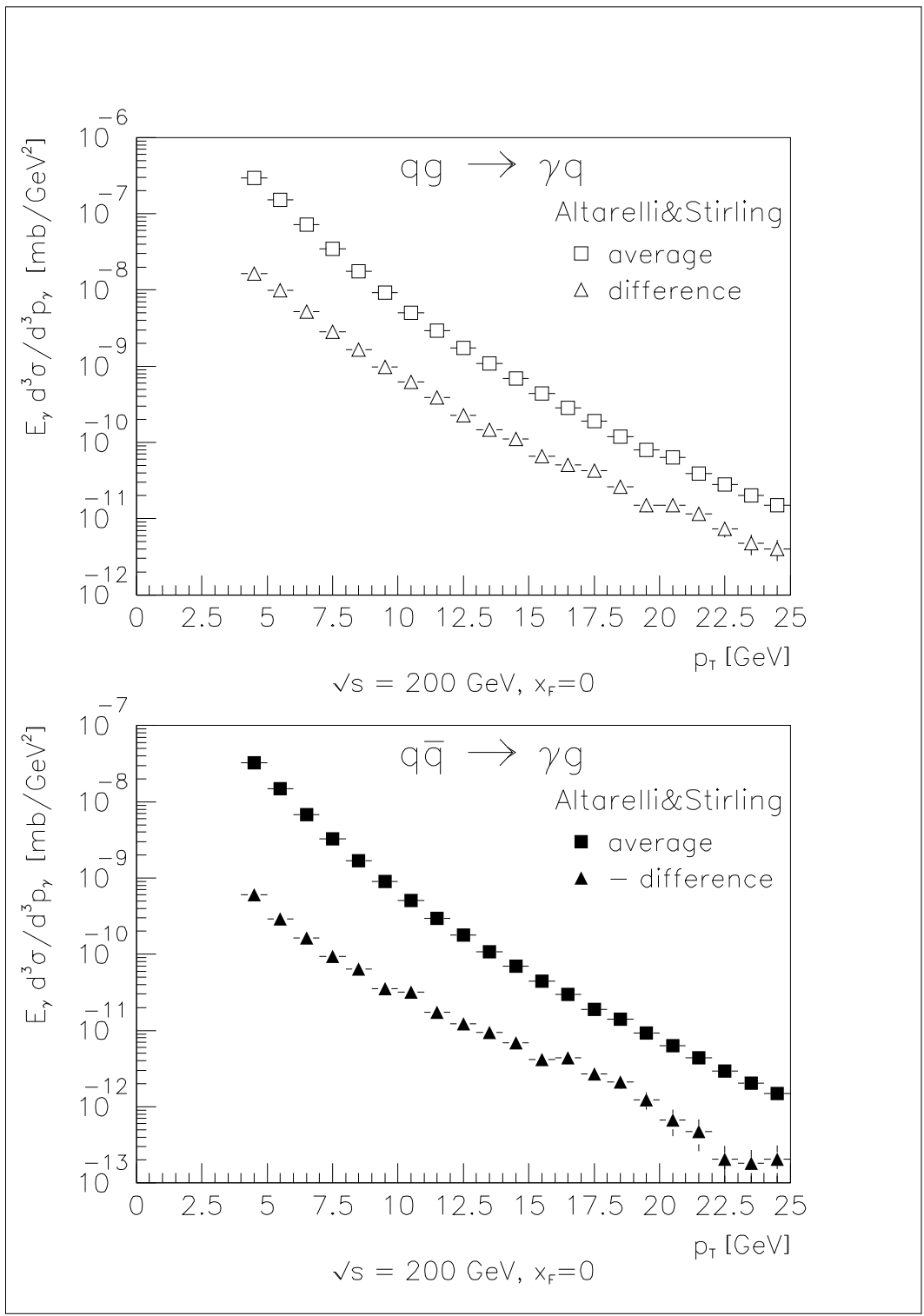}
  \caption[Wirkungsquerschnitt des Spinmittels und der Spindifferenz
f\"ur die Prompt-$\gamma$-Produktion]{\small\sl{\bf
Spin-average and spin-difference cross sections for prompt $\gamma$
production. (Note that in the lower plot the negative difference was
plotted.)
}\hfill\break
{\bf top:} The Compton process \ \ \ \
{\bf bottom:} The annihilation process}
  \label{fig3}
\end{figure}

\clearpage
\begin{figure}[htb]
  \centering
  \epsffile{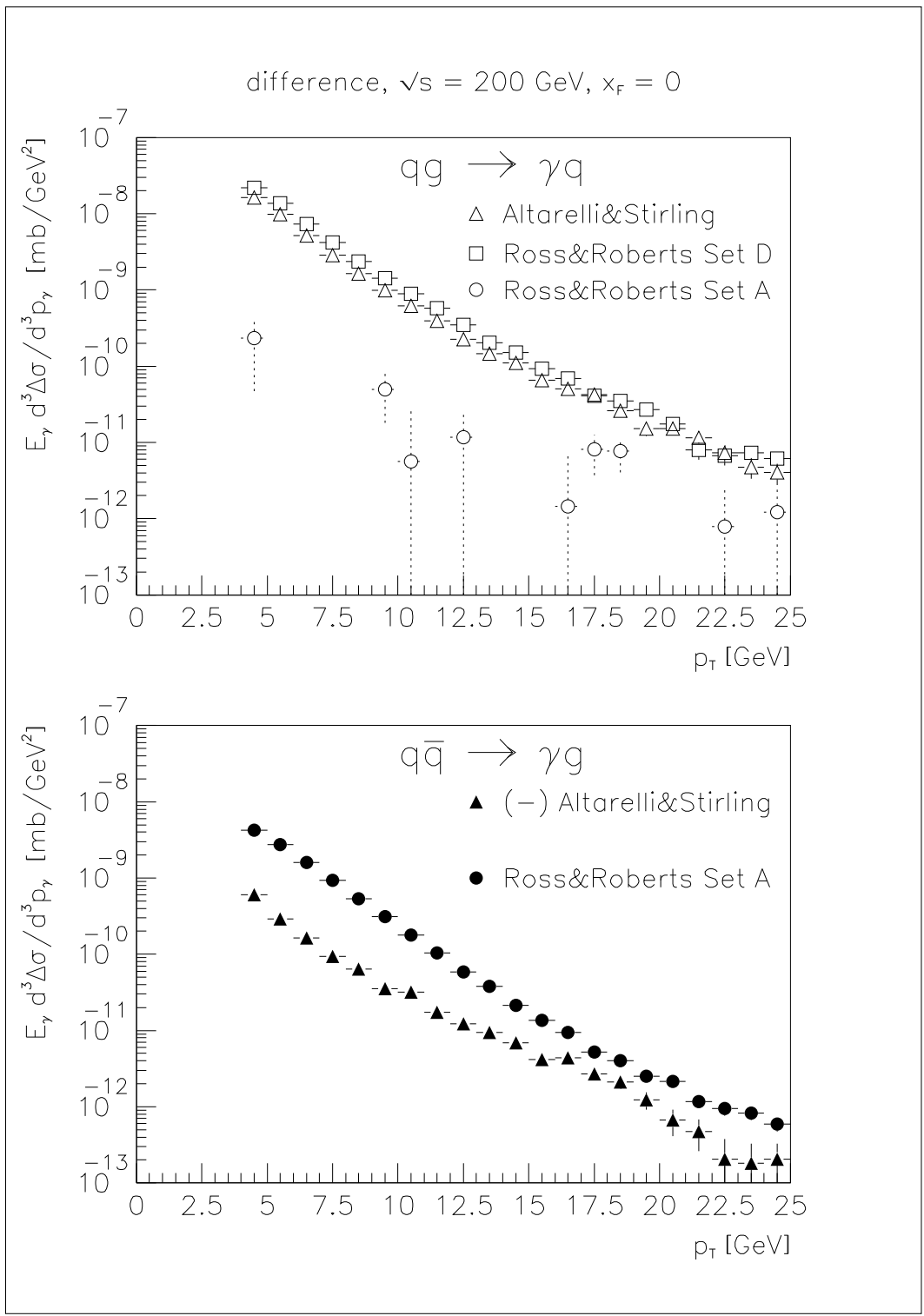}
  \caption[Wirkungsquerschnitt der Spindifferenz
f\"ur die Prompt-$\gamma$-Produktion f\"ur verschiedene
Parametrisierungen]{\small\sl{\bf
Spin-difference cross sections for prompt-$\gamma$
production for various parametrisations.
(Note that in the lower plot the negative difference was
plotted for the Altarelli\&Stirling parametrisation.)
}\hfill\break
{\bf top:} The Compton process \ \ \ \
{\bf bottom:} The annihilation process}
  \label{fig4}
\end{figure}

\clearpage
\begin{figure}[htb]
  \centering
  \epsffile{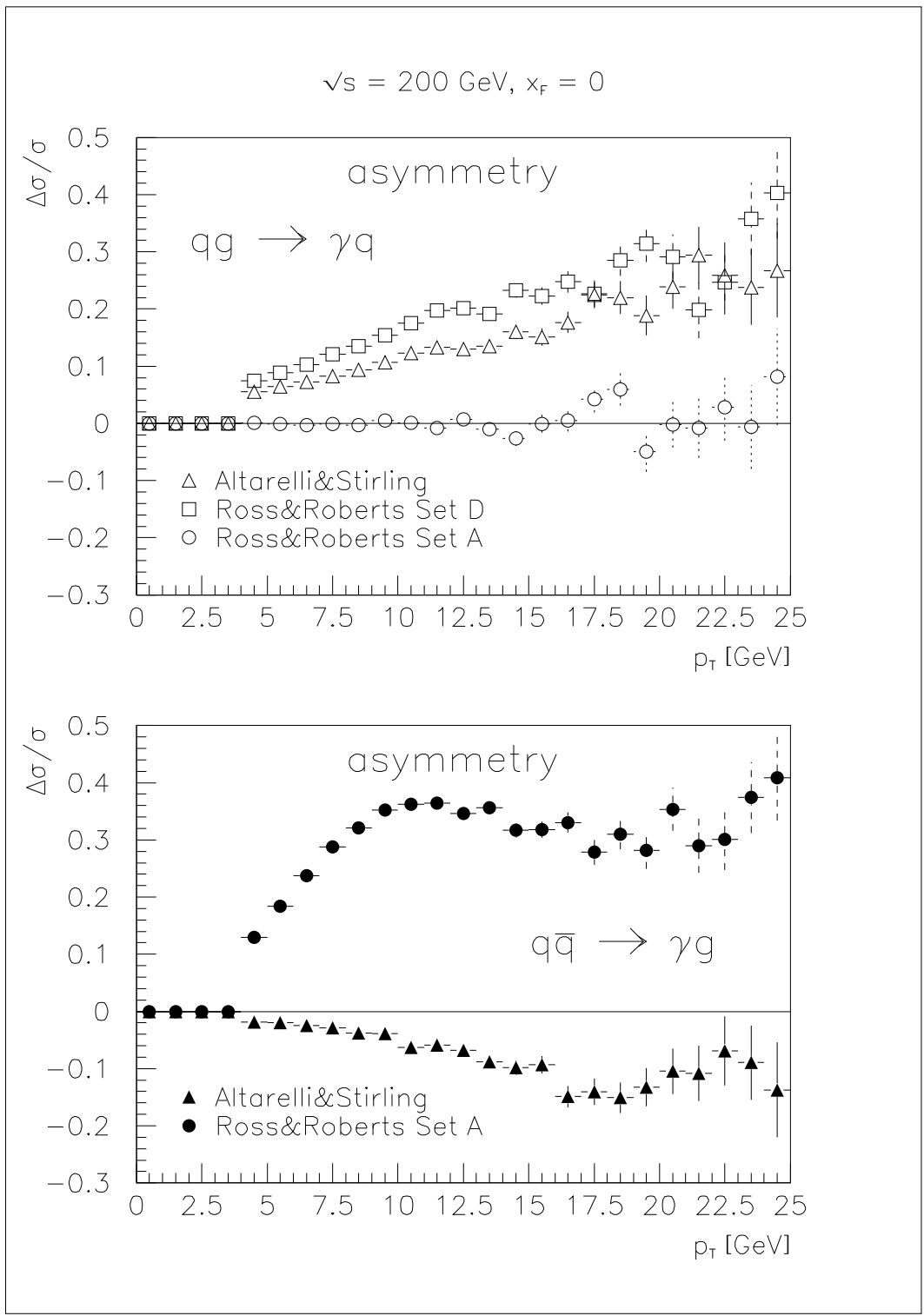}
  \caption[Die Asymmetrie f\"ur die Prompt-$\gamma$-Produktion f\"ur
verschiedene Parametrisierungen]{\small\sl{\bf
The asymmetry for prompt-$\gamma$ production for various
parametrisations
}\hfill\break
{\bf top:} The Compton process \ \ \ \
{\bf bottom:} The annihilation process}
  \label{fig5}
\end{figure}

\clearpage
\begin{figure}[htb]
  \centering
  \epsffile{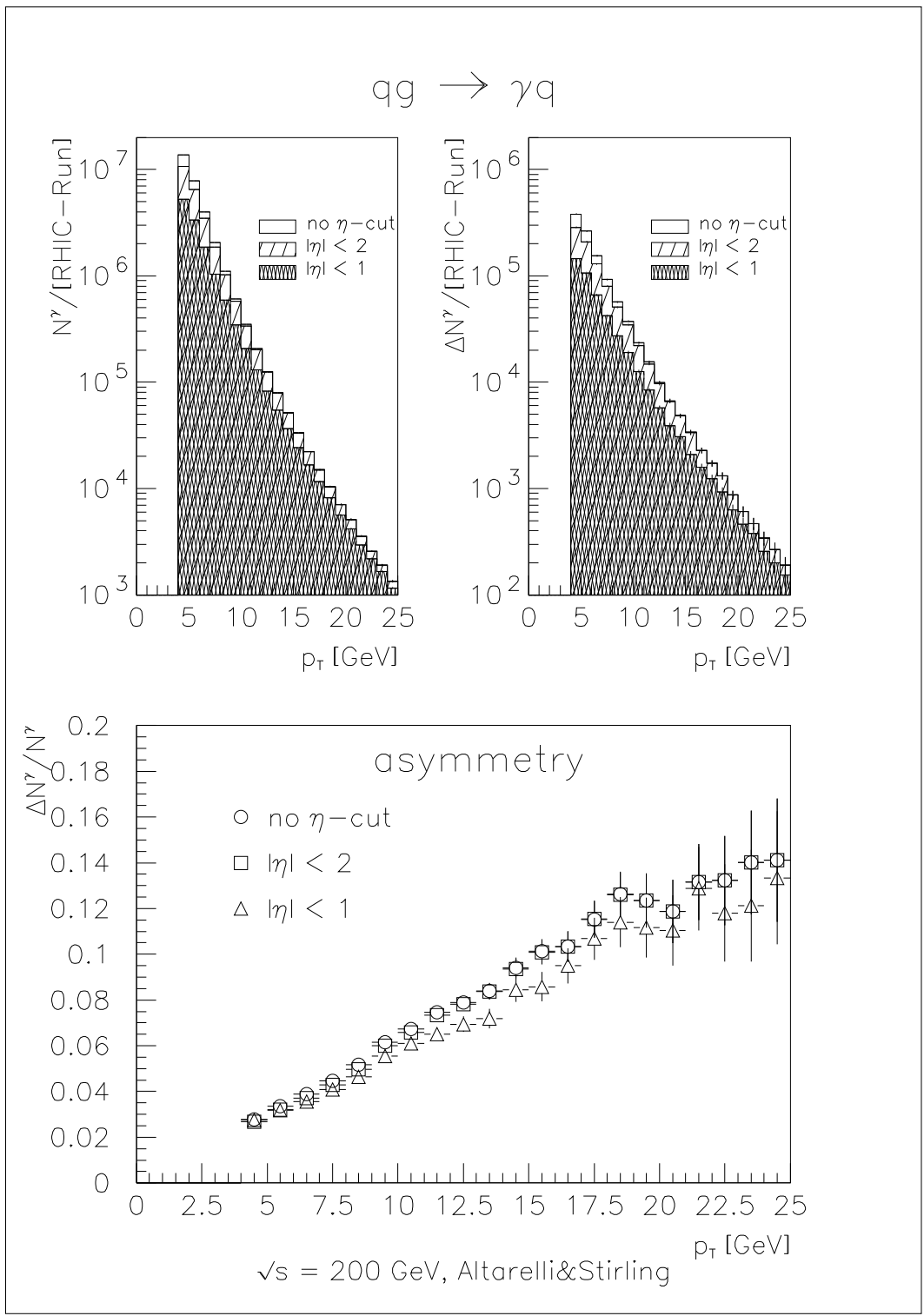}
  \caption[Einflu{\ss} des $\eta$-Cuts auf die
Prompt-$\gamma$-Produktion]{\small\sl
{\bf
Influence of the $\eta$-cut on prompt-$\gamma$ production
}\hfill\break
{\bf top left:} spin average \ \ \ \
{\bf top right:} spin difference \ \ \ \
{\bf bottom:} asymmetry}
  \label{fig6}
\end{figure}

\clearpage
\begin{figure}[htb]
  \centering
  \epsffile{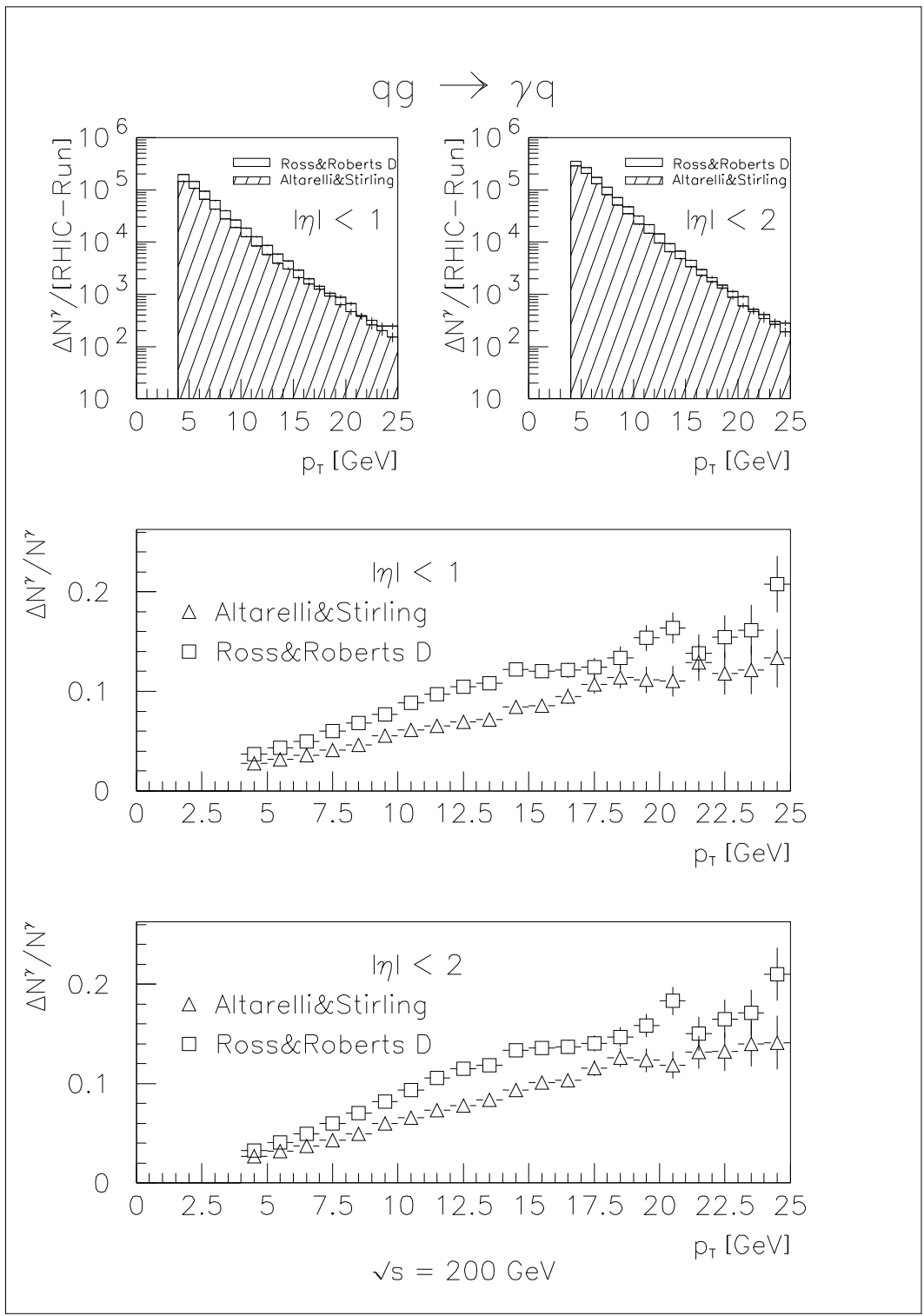}
  \caption[$\eta$-Cut bei verschiedenen
Parametrisierungen]{\small\sl{\bf
Consequences of the $\eta$-cut for two different parametrisations
}\hfill\break
{\bf top left:} spin difference, $\left|\eta\right|<1$ \ \ \ \ \
{\bf top right:} spin difference, $\left|\eta\right|<2$\hfill\break
{\bf middle:} asymmetry, $\left|\eta\right|<1$ \ \ \ \ \ \ \ \ \ \
{\bf bottom:} asymmetry, $\left|\eta\right|<2$
}
  \label{fig7}
\end{figure}

\clearpage
\begin{figure}[htb]
  \centering
  \epsffile{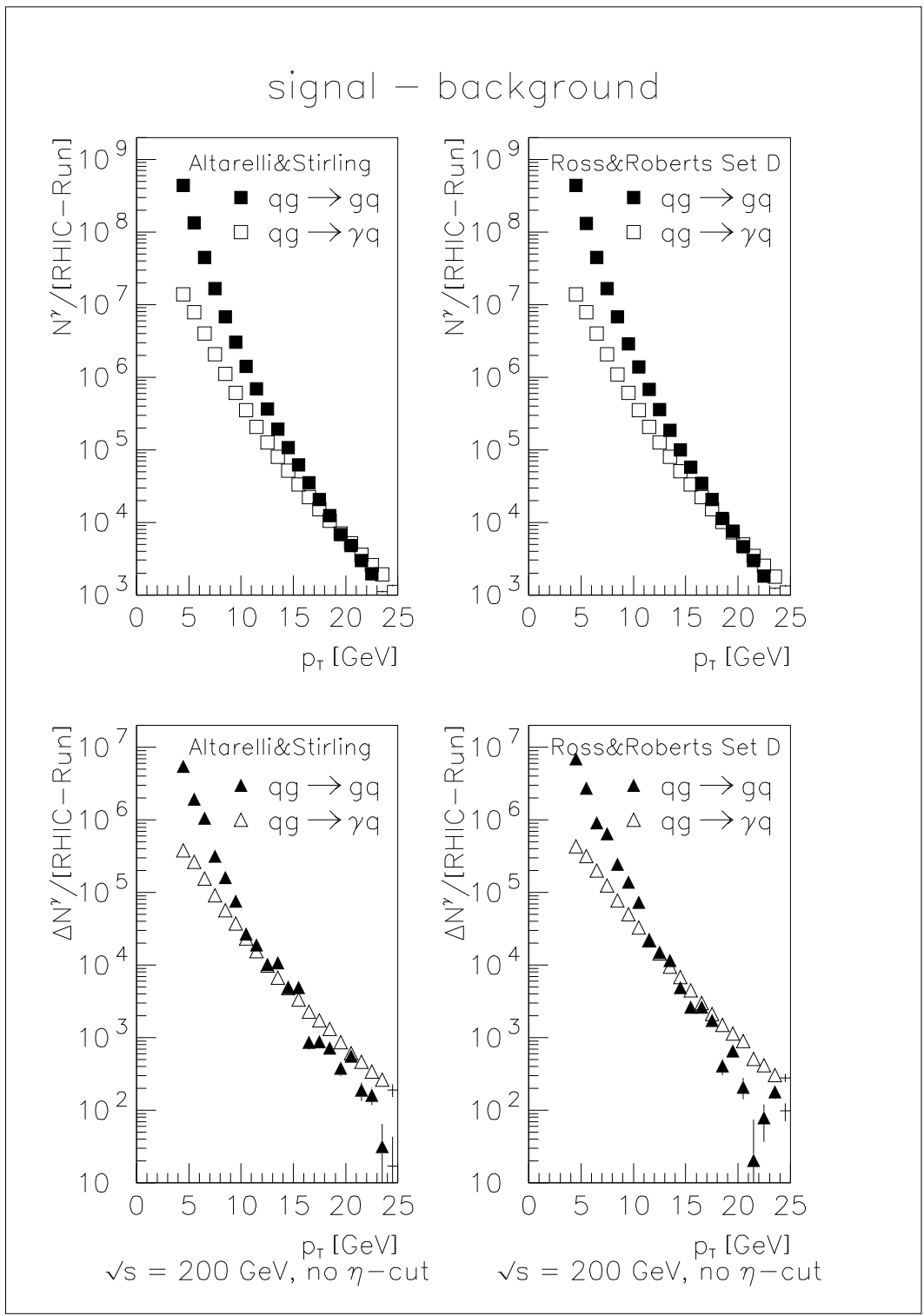}
  \caption[Vergleich von Signal und Untergrund]{\small\sl{\bf
Comparison of signal and background
}\hfill\break
{\bf top:} spin average \ \ \ \ \ \ \ \ \ \ \ {\bf left:} Altarelli\&Stirling
\hfill\break
{\bf bottom:} spin difference \ \ {\bf right:} Ross\&Roberts set D}
  \label{fig8}
\end{figure}

\clearpage
\begin{figure}[htb]
  \centering
  \epsffile{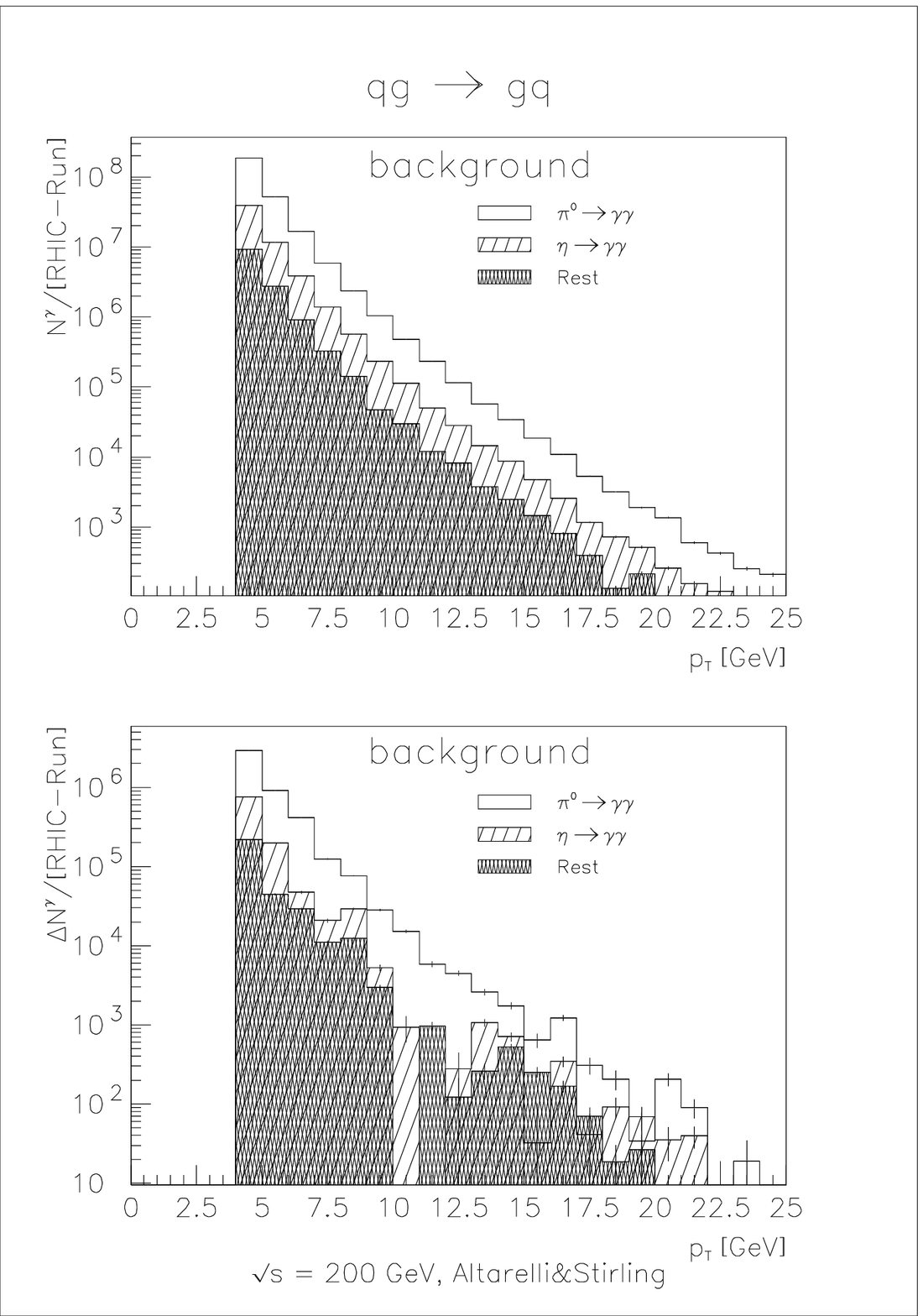}
  \caption[Die Zusammensetzung des Untergrunds f\"ur $qg\rightarrow
gq$]{\small\sl{\bf
Composition of the background for  $qg\rightarrow gq$}
\hfill\break
{\bf top:} spin average \ \ \ \
{\bf bottom:} spin difference}
  \label{fig9}
\end{figure}

\clearpage
\begin{figure}[htb]
  \centering
  \epsffile{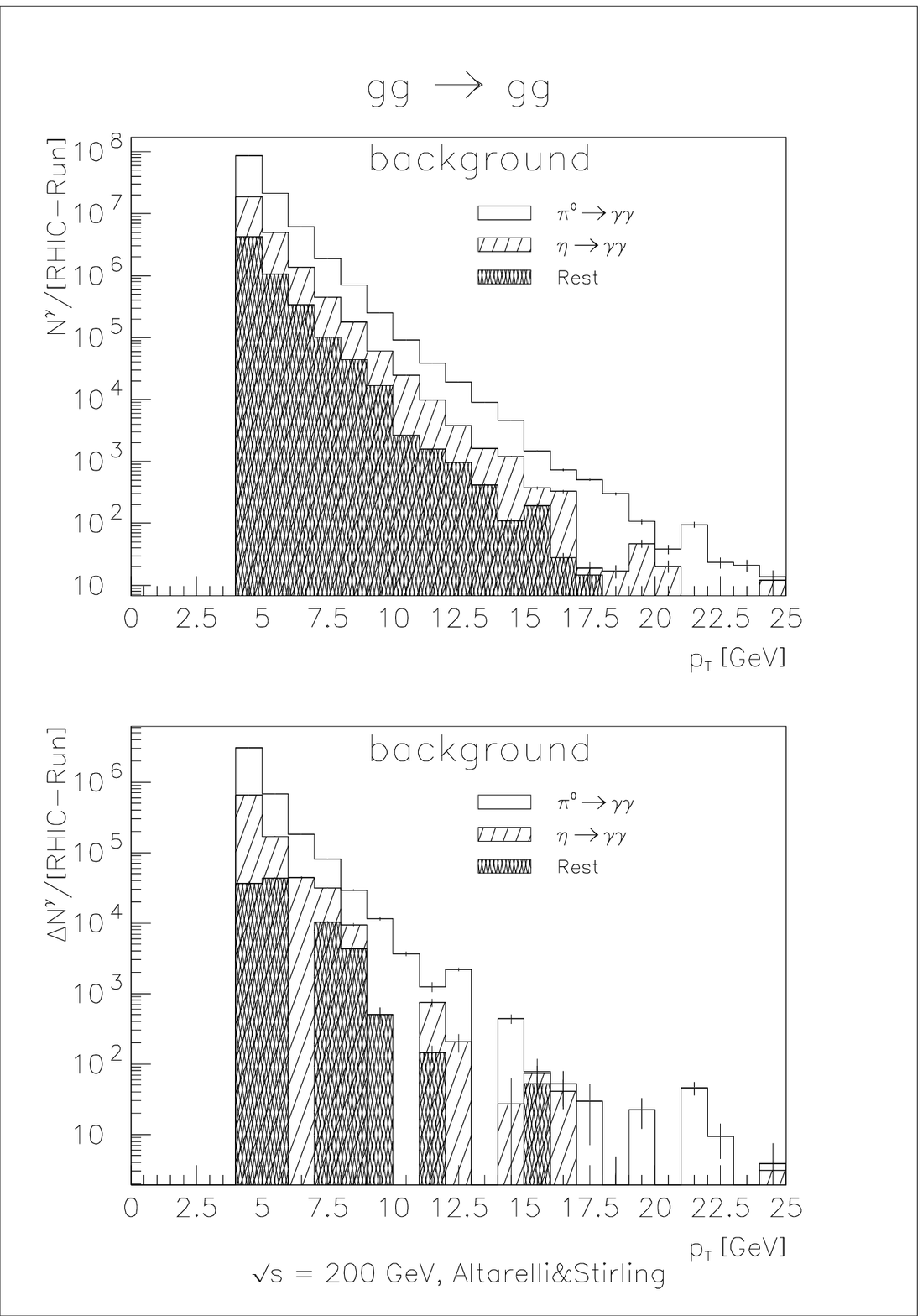}
  \caption[Die Zusammensetzung des Untergrunds f\"ur $qg\rightarrow
gq$]{\small\sl{\bf
Composition of the background for  $gg\rightarrow gg$}
\hfill\break
{\bf top:} spin average \ \ \ \
{\bf bottom:} spin difference}
  \label{fig10}
\end{figure}

\clearpage
\begin{figure}[htb]
  \centering
  \epsffile{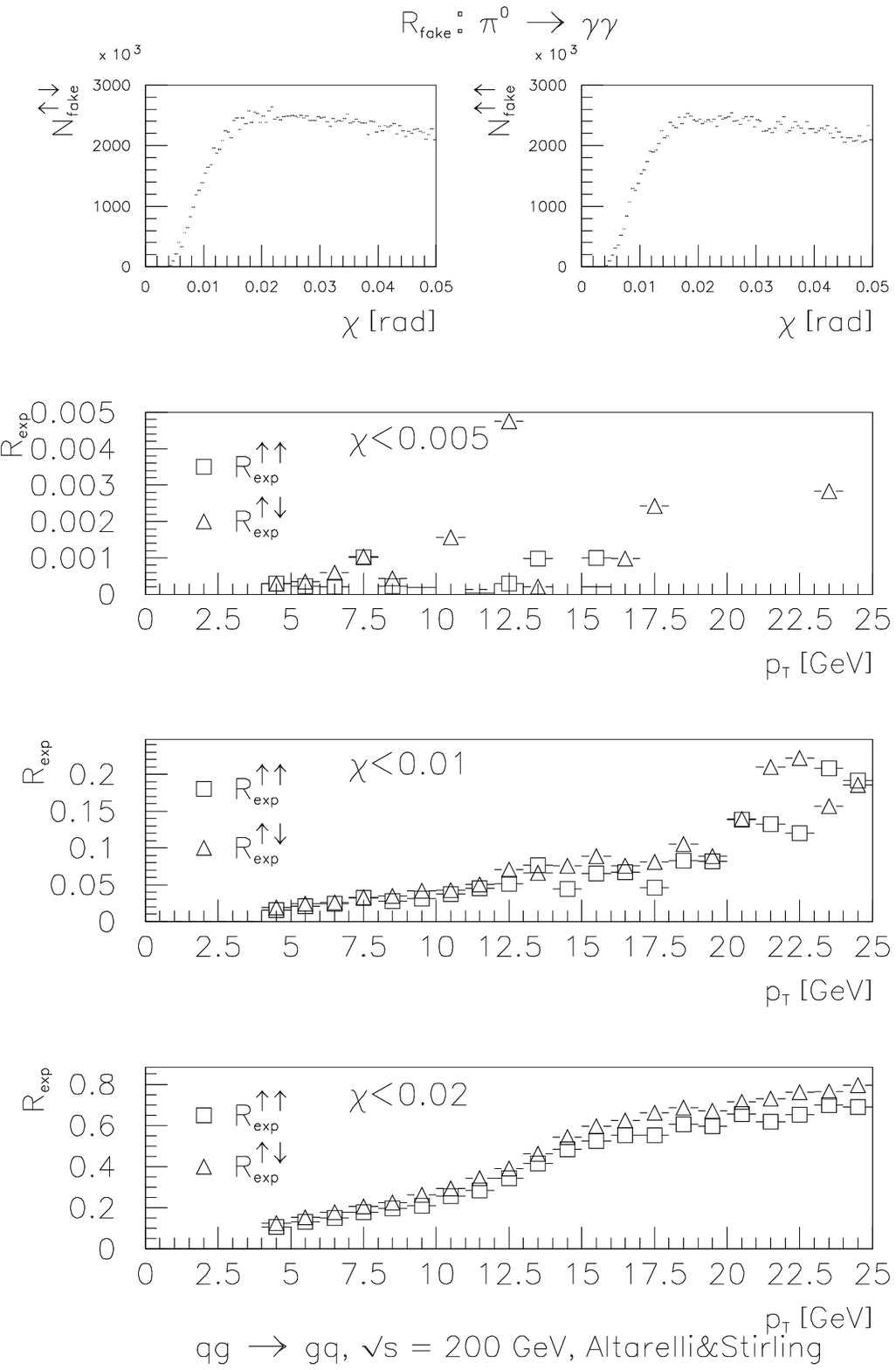}
  \caption[Fake-$\gamma$-Raten]{\small\sl{\bf
Fake-$\gamma$-rates}\hfill\break
{\bf 1. line:} distribution of opening angles for the photon pairs
\hfill\break
{\bf 2.-4. line:} fake-$\gamma$-rates for different resolutions}
  \label{fig11}
\end{figure}

\clearpage
\begin{figure}[ht]
  \centering
  \epsffile{
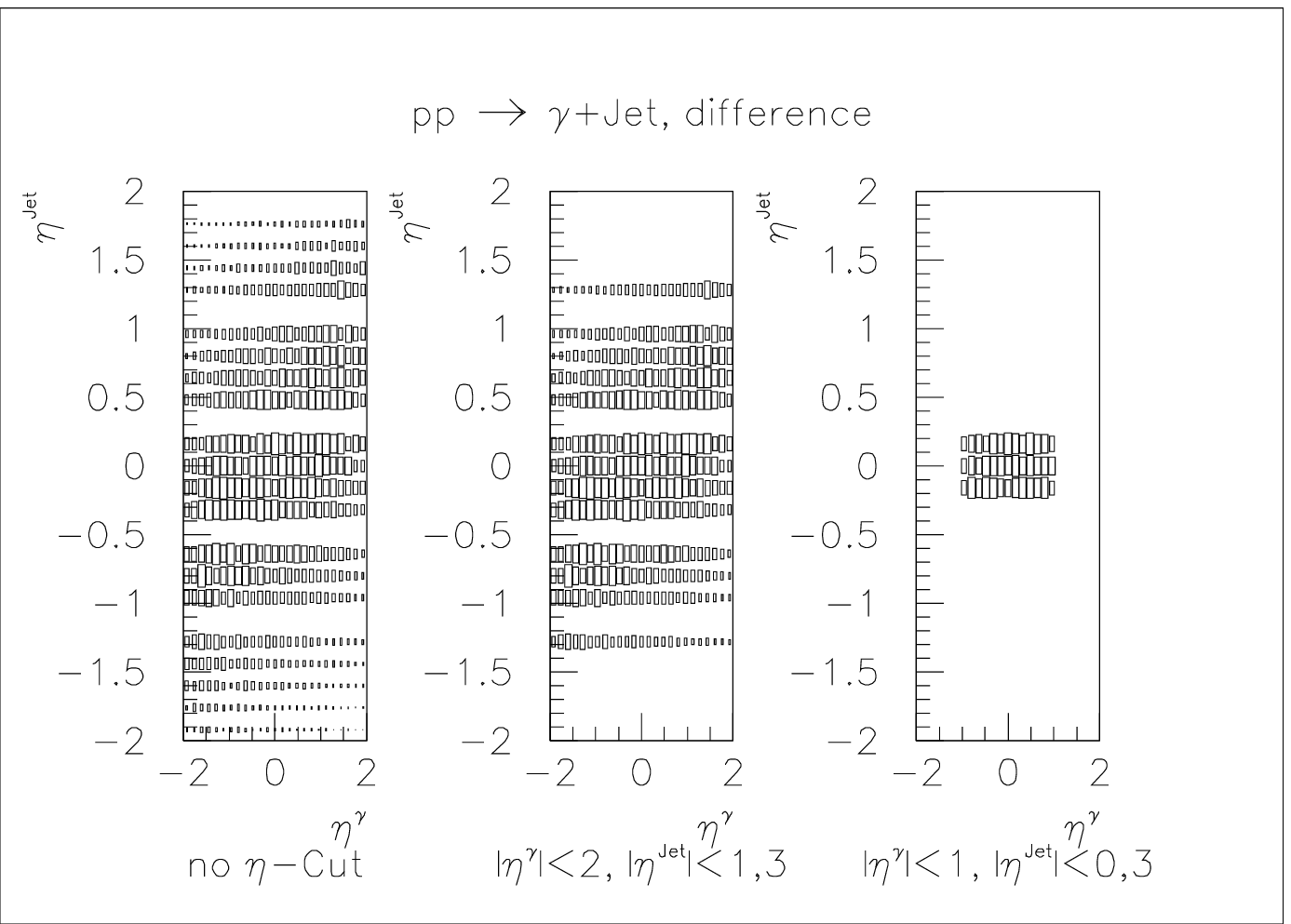}
  \caption[$\gamma$+Jet-Produktion f\"ur verschiedene $\eta$-Cuts]{\small\sl
{\bf
$\gamma$+jet-production for different $\eta$ cuts.
The areas of the rectangles are proportional to the logarithm of the
counts.}\hfill\break

{\bf left:} no $\eta$ cut \hfill\break
{\bf middle:} STAR with end caps \hfill\break
{\bf right:} STAR without end caps}
  \label{fig12}
\end{figure}

\clearpage
\begin{figure}[ht]
  \centering
  \epsffile{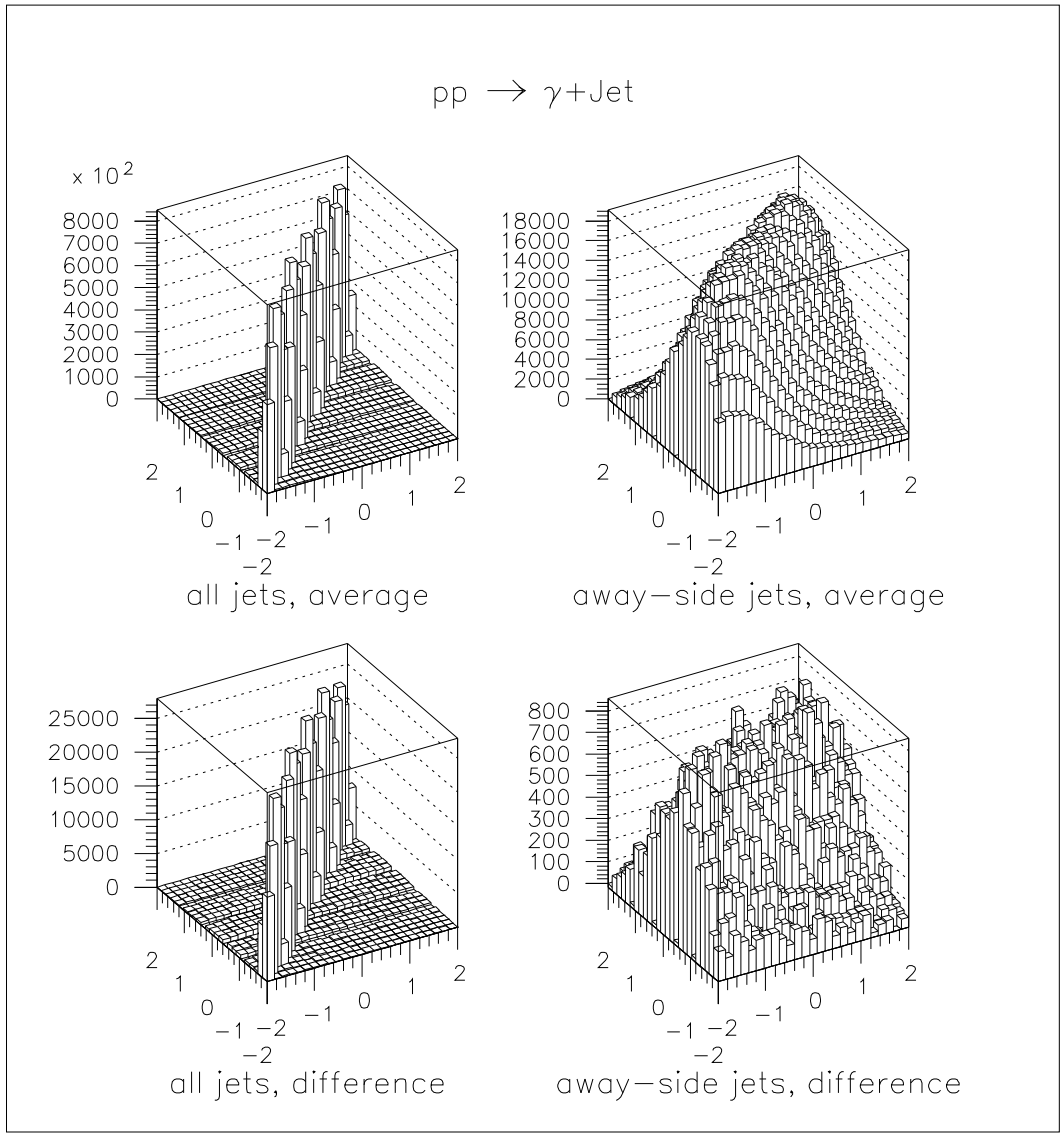}
  \caption[$\gamma$+Jet-Produktion: Alle Jets vs.\
Away-Side Jets]{\small\sl{\bf
Histogramms for the rapidity distribution of jet events. The photon
rapidity $\eta^{\gamma}$ is plotted to the
backward-right, the jet-rapidity $\eta^{\rm Jet}$ is
plotted to the backward-left, both in the range from -2 to 2. Spin
average and spin difference are compared for all jets and away-side
jets.
}\hfill\break
{\bf top:} spin average \ \ \ \,
{\bf bottom:} spin difference \hfill\break
{\bf left:} all jets \ \ \ \ \ \ \ \ \ \ \ \ \ {\bf right:} away-side jets
only}
  \label{fig13}
\end{figure}

\clearpage
\begin{figure}[ht]
  \centering
  \epsffile{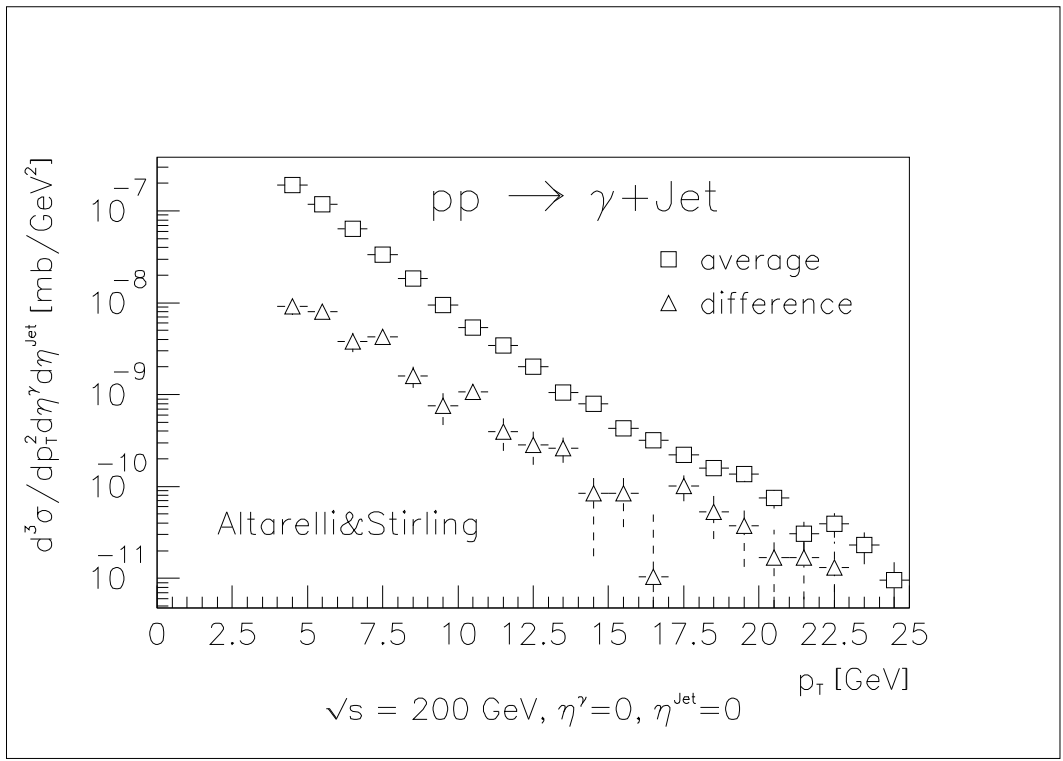}
  \caption[Prompt-$\gamma$-Produktion und Away-Side Jet:
Altarelli\&Stirling]{\small\sl{\bf
Cross section for prompt-$\gamma$ production with an away-side jet
for the Altarelli\&Stirling parametrisation}}
  \label{fig14}
\end{figure}

\clearpage
\begin{figure}[ht]
  \centering
  \epsffile{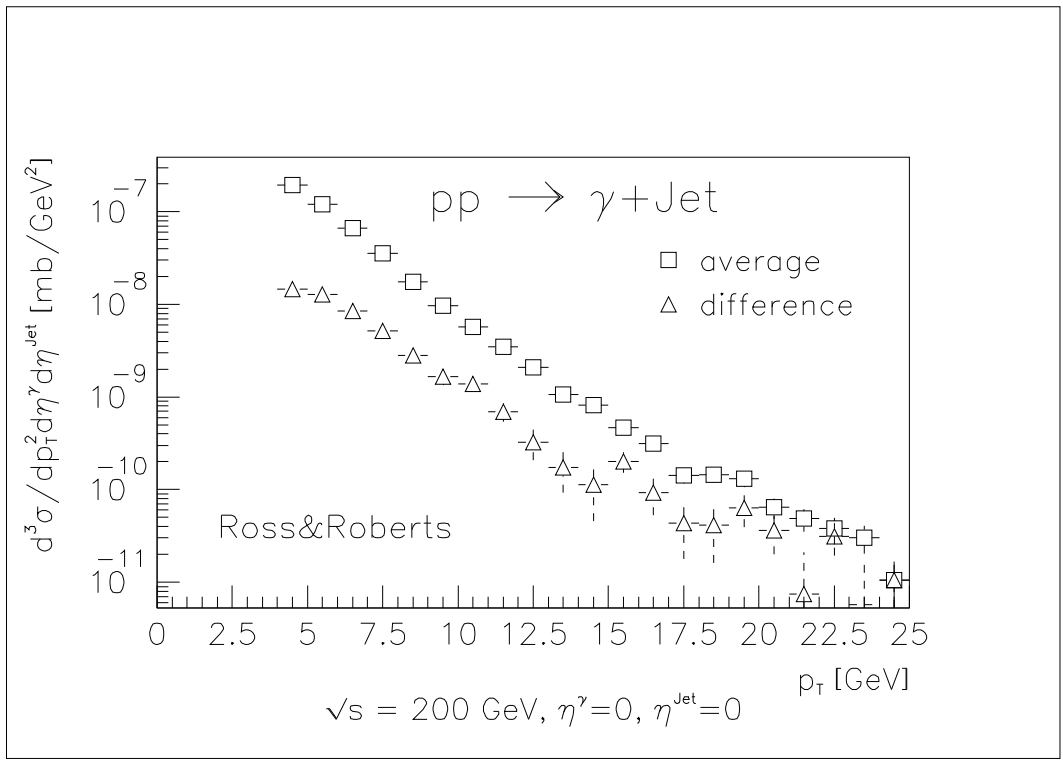}
  \caption[Prompt-$\gamma$-Produktion und Away-Side Jet:
Ross\&Roberts]{\small\sl{\bf
Cross section for prompt-$\gamma$ production with an away-side jet
for the Ross\&Roberts parametrisation}}
  \label{fig15}
\end{figure}

\clearpage
\begin{figure}[ht]
  \centering
  \epsffile{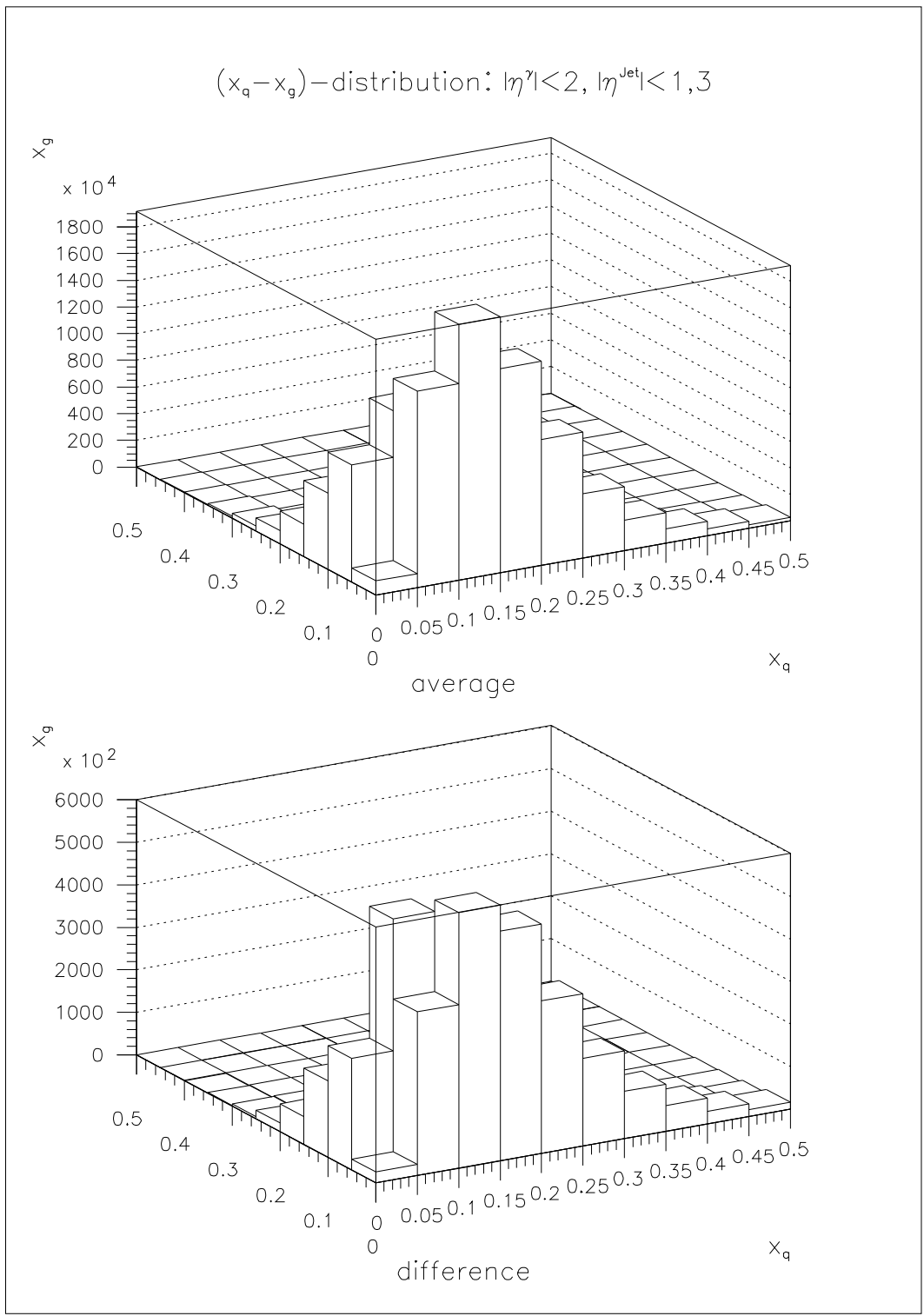}
  \caption[(Simulierte $x_q$-$x_g$)-Verteilung f\"ur den STAR-Detektor mit
Endkappen]{\small\sl{\bf
Simulated ($x_q$-$x_g$) distribution for the STAR detector with
end caps. $x_q$ is plotted to the right, $x_g$ to the left backward.
}\hfill\break
{\bf top:} spin average \ \ \ \
{\bf bottom:} spin difference}
  \label{fig16}
\end{figure}

\clearpage
\begin{figure}[ht]
  \centering
  \epsffile{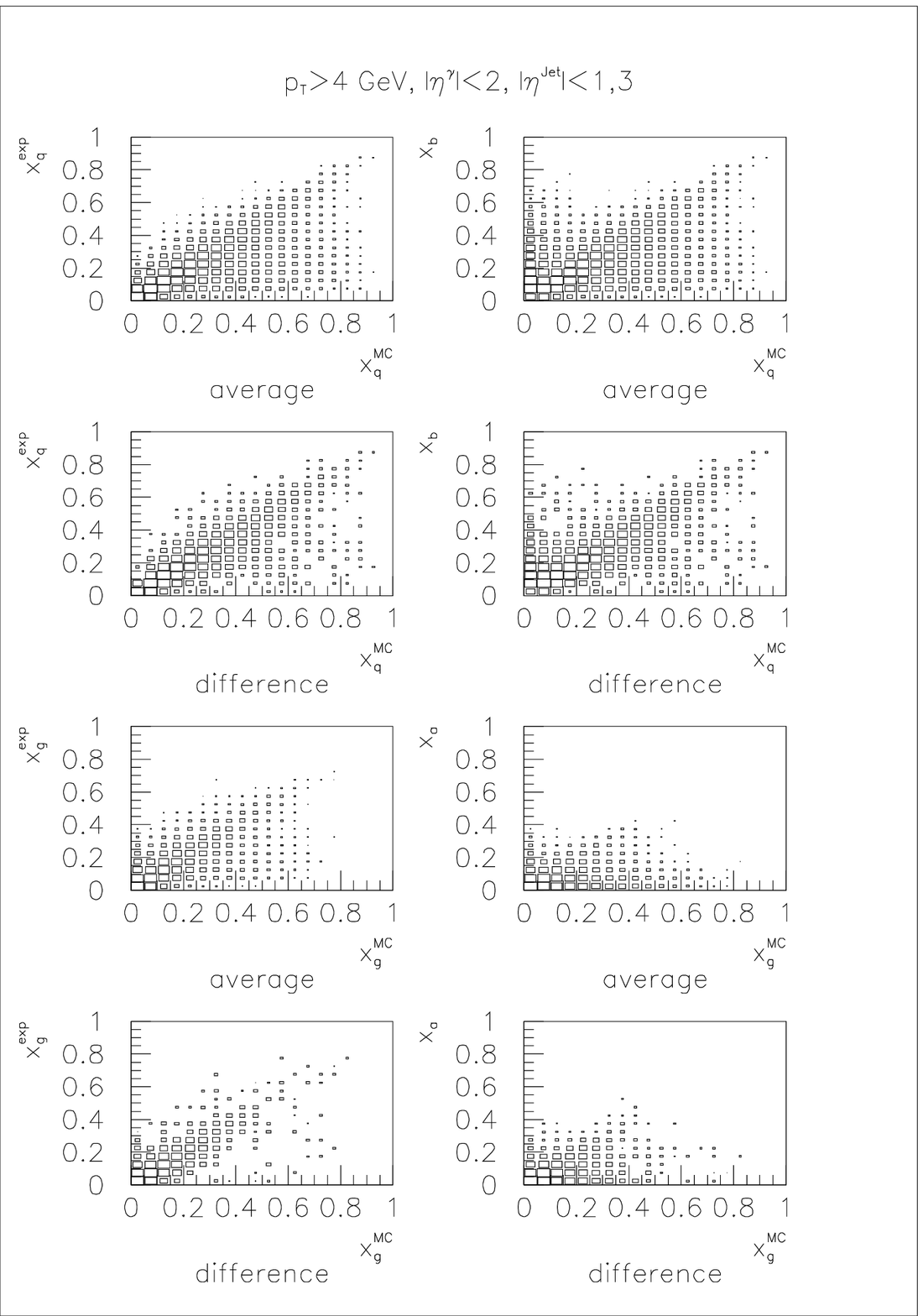}
  \caption[$x$-Rekonstruktion f\"ur Quark und Gluon. Experiment vs.\
Simulation]{\small\sl{\bf
$x$ reconstruction for quarks and gluons,
idealy $x^{\rm MC}_q$ should be strictly correlated with $x^{\rm exp}_q$
and
$x_b$ and
$x^{\rm MC}_g$ with $x^{\rm exp}_g$ and
$x_a$. The areas of the rectangles are proportional to the logarithm of the
counts.}\hfill\break
{\bf 1.+2. line:} quark \ \ \ \ \ \ \ \ \ \ \, {\bf 3.+4.
line:} gluons \ \ \ \ \ \ \ \ \ \ \ \ {\bf left:}
$x_{MC}$ vs.\ $x_{exp}$ \hfill\break
{\bf 1.+3. line:} spin average \ \ {\bf 2.+4. line:}
spin difference \
{\bf right:} $x_{a/b}$ vs.\ $x_{g/q}$\hfill\break}
  \label{fig17}
\end{figure}

\clearpage
\begin{figure}[ht]
  \centering
  \epsffile{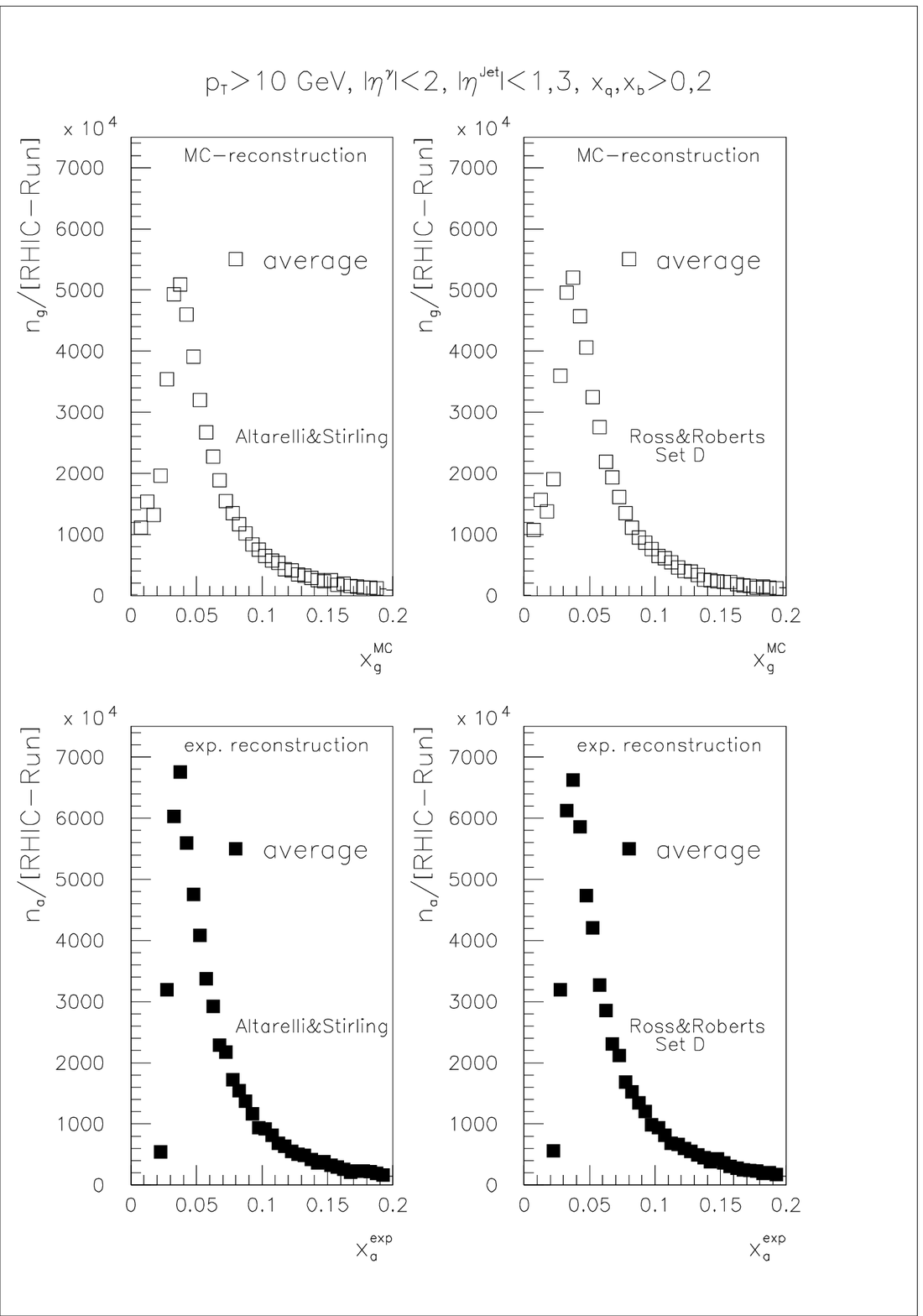}
  \caption[Rekonstruierte $x_g$-Verteilung f\"ur das Spinmittel, MC
vs.\ Experiment]{\small\sl{\bf
Reconstructed spin averaged $x_g$-distributions. MC-reconstraction
implies that cuts on $x_q$ were applied while for `experimental
reconstruction' these cuts were imposed on $x_b.$
}\hfill\break
{\bf top:} MC reconstruction \ \ \ \ \ \ \ \ \ \ \ \ \ \ \ \ \ \ \ \ \ \ \ \
\ \ \ \ \ \ \ \ \
{\bf left:} Altarelli\&Stirling\hfill\break
{\bf bottom:} Exp.\ reconstruction \ \ \ \ \ \ \ \ \ \ \ \ \ \ \ \ \ \ \ \ \ \
\ \ \ \ \
{\bf rechts:} Ross\&Roberts set D\hfill\break}
  \label{fig18}
\end{figure}

\clearpage
\begin{figure}[ht]
  \centering
  \epsffile{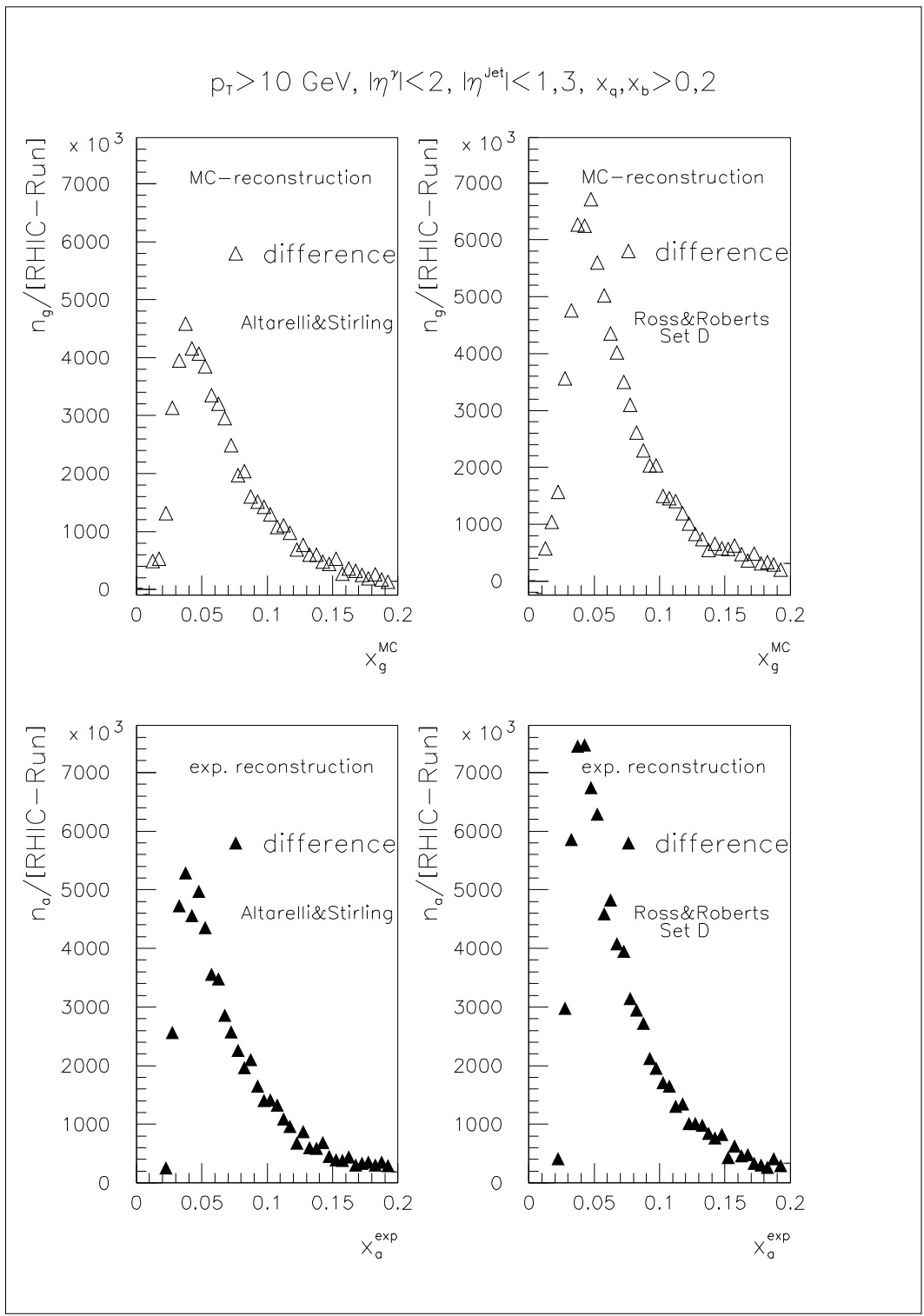}
  \caption[Rekonstruierte $x_g$-Verteilung f\"ur die
Spindifferenz, MC vs.\ Experiment]{\small\sl{\bf
Same as figure 18 for the spin difference}\hfill\break}
  \label{fig19}
\end{figure}

\clearpage
\begin{figure}[htb]
  \centering
  \epsffile{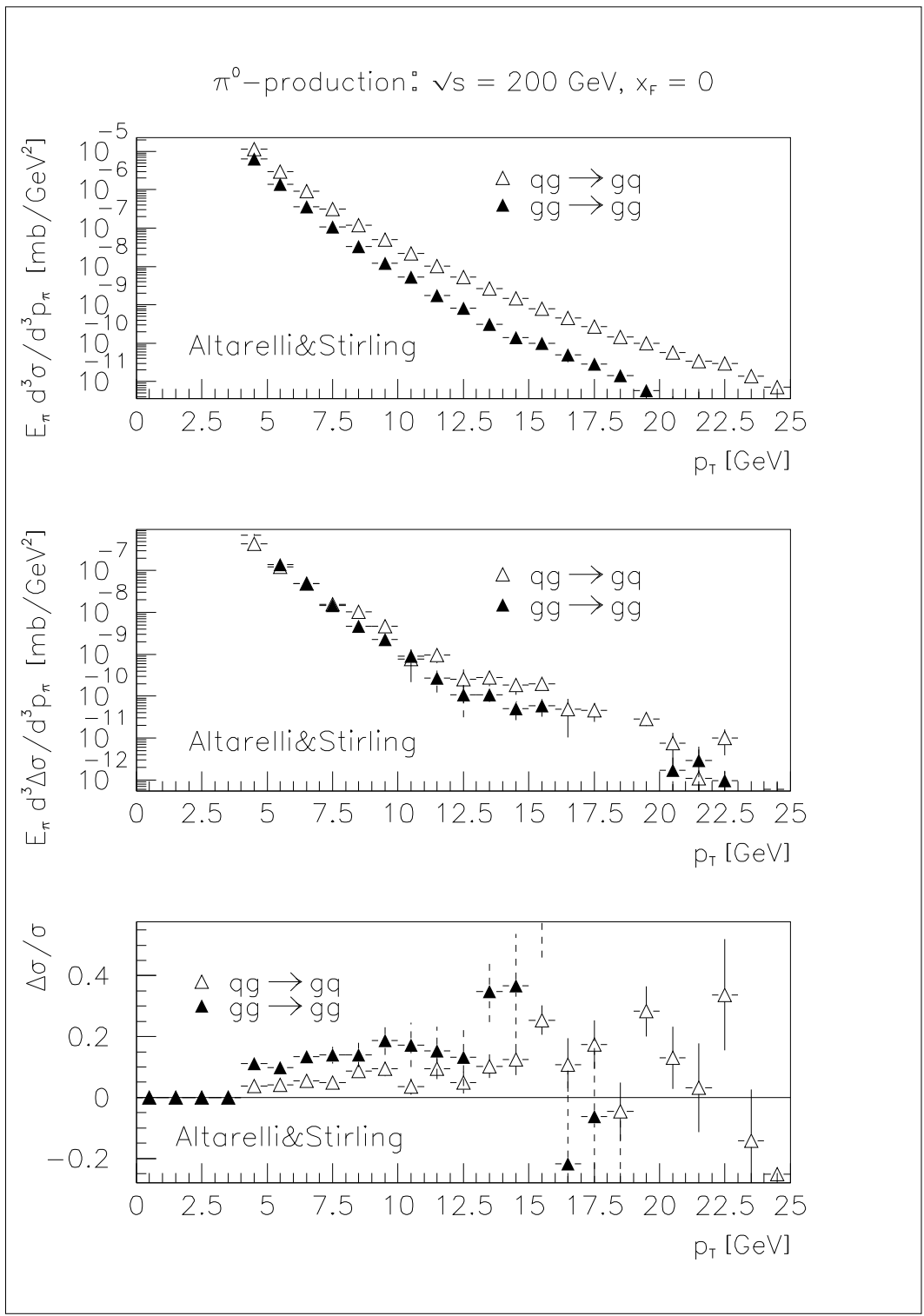}
  \caption[$\pi^0$-Produktion: $qg\rightarrow gq$ vs.\
$gg\rightarrow gg$]{\small\sl{\bf
$\pi^0$ production to the hard processes $qg\rightarrow gq$ and
\ $gg\rightarrow gg$}\hfill\break
{\bf top:} spin average\ \ \ \
{\bf middle:} spin difference \ \ \ \
{\bf bottom:} asymmetry}
  \label{fig20}
\end{figure}

\clearpage
\begin{figure}[htb]
  \centering
 ~
  \epsffile{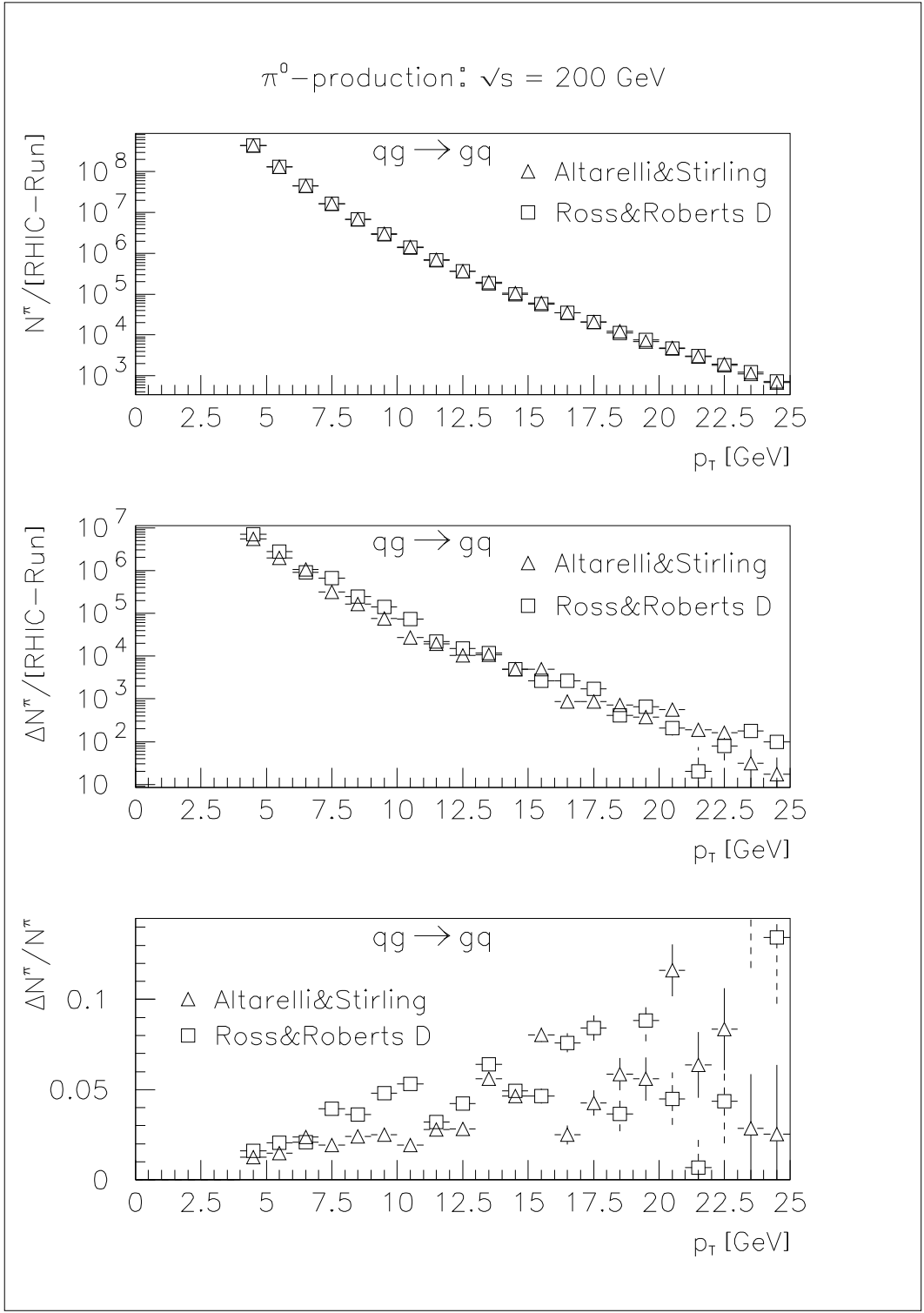}
  \caption[$\pi^0$-Produktion: Altarelli\&Stirling vs.\ Ross\&Roberts
Satz D]{\small\sl{\bf
$\pi^0$ production: comparison of the results for the Altarelli\&Stirling
and the \ Ross\&Roberts D parametrisation}\hfill\break
{\bf top:} spin average \ \ \ \
{\bf middle:} spin difference \ \ \ \
{\bf bottom:} asymmetry}
  \label{fig21}
\end{figure}
\end{document}